\documentclass[aps,pre,twocolumn,amsmath,amssymb,floatfix,amsfonts,showkeys,superscriptaddress]{revtex4}
\usepackage{epsfig,color}
\usepackage{times}
\usepackage{amsfonts}
\usepackage{amssymb}
\usepackage{amsmath}
\usepackage{graphicx}
\usepackage{hyperref}
\usepackage{color} 
\usepackage{verbatim}
\begin{document}
\title{Eco-Evolutionary Dynamics of a Population with Randomly Switching Carrying Capacity}

\author{Karl Wienand}
\affiliation{Arnold Sommerfeld Center for Theoretical Physics, Department of Physics, 
Ludwig-Maximilians-Universit\"at M\"unchen, Theresienstrasse 37, 80333 M\"unchen, Germany}

\author{Erwin Frey}
\affiliation{Arnold Sommerfeld Center for Theoretical Physics, Department of Physics, 
Ludwig-Maximilians-Universit\"at M\"unchen, Theresienstrasse 37, 80333 M\"unchen, Germany}

\author{Mauro Mobilia}
\email{m.mobilia@leeds.ac.uk}
\affiliation{Department of Applied Mathematics, School of Mathematics, University of Leeds, Leeds LS2 9JT, U.K.}

\begin{abstract}
Environmental variability greatly influences the eco-evolutionary dynamics of a population, i.e. it affects how its size and composition evolve.
Here, we study a well-mixed population of finite and fluctuating size whose growth is limited by a randomly switching carrying capacity.
This models the environmental fluctuations between states of resources abundance and scarcity.
The population consists of two strains, one growing slightly faster than the other, competing under two scenarios: one in which competition is solely for resources, and one in
which the slow (``cooperating'') strain produces a public good that benefits also the fast (``freeriding'') strain.
We investigate how the coupling of demographic and environmental (external) noise affects the population's 
eco-evolutionary dynamics.
By analytical and computational means, we study the correlations between the population size and its composition, 
and discuss the social-dilemma-like  ``eco-evolutionary game'' characterizing the public good production.
We determine in what conditions it is best to produce a public good; 
when cooperating is beneficial but outcompeted by freeriding, 
and when the public good production is detrimental for cooperators.
Within a  linear noise approximation to populations of varying size,
we also accurately analyze the coupled effects of demographic and environmental noise on the size distribution.
\end{abstract}


\maketitle

\section{Introduction}
\label{intro}

\begin{figure*}[htb]
\centerline{
\includegraphics[width=0.585\linewidth]{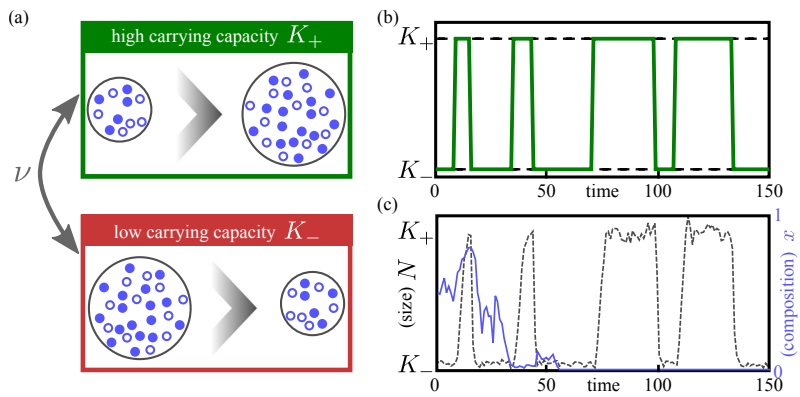}
}
\caption{
(a) Cartoon of the eco-evolutionary dynamics of the model: the population consists of strains $S$ ($\circ$) and 
$F$ ($\bullet$), subject to $K(t)\in \{K_-,K_+\}$ that randomly switches, see (\ref{eq:K(t)}).
After each switch of $K(t)$, $N$ and $x$ change: following a $K_-$ to $K_+$ switch, $N$ increases and the 
intensity of the internal noise decreases; the opposite occurs following a $K_+$ to $K_-$ switch. 
(b) Typical random switching of $K(t)$  according to (\ref{eq:K(t)}).
(c) Sample paths of $N(t)$ (gray, dashed line) and $x(t)$ (blue, solid line), corresponding to the switching portrayed in (b).
We notice that $x$ evolves much slower than $N$, see text. Parameters are $(s,\nu, K_+, K_-,b)=(0.02,0.1, 450, 50, 0)$.
}
\label{fig:Fig1} 
\end{figure*}

The fate of populations is affected by a number of endlessly changing environmental conditions such as the presence of toxins, resources abundance, temperature, light, etc. \cite{Morley83,Fux05}.
In the absence of detailed knowledge of how external factors vary, they are modeled as external noise (EN) shaping the randomly changing environment in which species evolve.
The impact of fluctuating environments on population dynamics  has been studied in a number of systems \cite{May73,Karlin74,He10,Tauber13,Assaf13,AMR13,Ashcroft14,Melbinger15,Hufton16,Hidalgo17,Kussell05b,Shnerb17}, and several evolutionary responses to exogenous changes have
 been analyzed~\cite{Chesson81,Kussell05,Acar08,Loreau08,Beaumont09,Visco10,Xue17}.
In finite populations, internal noise is another important form of randomness, yielding  demographic fluctuations of stronger intensity in small populations than in large ones.
Internal noise (IN) is responsible for fixation~\cite{Kimura,Blythe07,Nowak} (when one species takes 
over and others are wiped out) and thus plays an important role in the evolution of a population's composition.
Ecological and evolutionary dynamics are often coupled, through an interdependent evolution of the population size and composition \cite{Roughgarden79,Leibler09,Pelletier09,Harrington14,Melbinger2010,Cremer2011,Melbinger2015a}.
As a consequence, environmental variability may affect the population size and hence the demographic fluctuations intensity, thus coupling EN and IN.
The interdependence of environmental noise and demographic fluctuations is particularly relevant for microbial 
communities, whose properties greatly depend on the population size and on the environment~\cite{Morley83,Fux05}.
These populations often experience sudden environmental changes that can drastically affect their size, {\it e.g.}
by leading to \emph{population bottlenecks} under which the colony of reduced size is  more prone to fluctuations \cite{Wahl02,Patwas09,Brockhurst07a,Brockhurst07b}.
The coupling between the different forms of randomness therefore generates feedback loops between socio-biological interactions and the environment
\cite{Wahl02,Patwas09,Wienand15,Wienand18}, which results in fascinating eco-evolutionary phenomena such as cooperative behavior.
For instance, experiments on {\it Pseudomonas fluorescens} showed that the formation and sudden collapse of biofilms promotes the evolution of cooperation \cite{Brockhurst07a,Brockhurst07b,Rainey03}. 
In most studies, however, EN and IN are treated as uncoupled \cite{Karlin74,He10,Tauber13,Assaf13,AMR13,Ashcroft14,Kussell05b,Melbinger15,Hufton16,Hidalgo17,Shnerb17}. 
 
Recently, we introduced a model describing a fluctuating population---consisting of a fast strain competing with a slow (cooperating) species, that can produce a public good---evolving under a randomly switching carrying capacity \cite{KEM1}.
In this model, \emph{demographic fluctuations are coupled to EN}, resulting in a significant influence on the species fixation probability and leading to noise-induced transitions of the population size.
In the context of the eco-evolutionary dynamics of this model, 
here we introduce the theoretical concept of ``eco-evolutionary game'' to characterize the emergence of cooperation in populations of fluctuating size. 
We study the correlations between the population size and its composition and show that a social dilemma of sorts arises: while the public good production increases the overall expected population size, it also
lowers the survival probability of cooperators.
In the biologically-inspired setting of a metapopulation of non-interacting communities of varying size,
we measure the success of each species in the eco-evolutionary game in terms of its expected long-term number 
of individuals. We thus determine the circumstances under which public good production (cooperation) is detrimental or beneficial to 
cooperators, and find the conditions in which it is best to produce the public good.
Furthermore, we have devised a linear noise approximation that allows us to
accurately characterize the population size distribution and noise-induced transitions in a population whose size fluctuates under the joint effect of coupled demographic 
and environmental noise.

The next two sections establish our approach:
In section \ref{Model}, we introduce our stochastic model; in section \ref{sec:PDMP-Moran-fixation}, we outline the properties of the 
fitness-dependent Moran model and piecewise deterministic Markov processes associated with the model, and review how to combine these to compute the species fixation probabilities.
In the following two sections, we present our main results: Section \ref{sec:public-good} is dedicated to the correlations between the population size and its composition, and 
to the discussion of the emergence of cooperative behavior along with an ``eco-evolutionary game'' in a population of fluctuation size;
in section \ref{INEN}, we study the population size distribution within a linear noise approximation.
Our conclusions are presented in Sec. \ref{final}.
Additional information is provided in the Supplementary Material (SM) \cite{Supp}.

\section{Model}
\label{Model}

As in our recent work \cite{KEM1}, we consider a well-mixed population of fluctuating size $N(t)=N_S(t)+N_F(t)$, 
consisting of $N_S$ individuals of species $S$ and $N_F$ of species, or strain, $F$~\cite{Strains}.
The fast-growing strain $F$ has fitness $f_F=1$, whereas the slow-growing  strain $S$ has a slightly lower fitness $f_S=1-s$, 
where $0<s\ll1$ denotes the (weak) selection intensity. At time $t$ the fraction of $S$ individuals in the population is $x(t)=N_S(t)/N(t)$ and 
the average population  fitness is $\bar f=x f_S+(1-x)f_F=1-sx=1+{\cal O}(1)$. Here, the evolution of the
population size $N(t)$ is coupled to the internal composition $x(t)$ by 
a global growth rate $g(x)$, and its growth is limited by a logistic death rate 
$N/K(t)$~\cite{Roughgarden79,Melbinger2010,Cremer2011,KEM1}.
The carrying capacity $K(t)$ is a measure of the population size that can be supported, and is here assumed
to vary in time, see below.
We specifically focus on two important forms of global growth rates:
(i) the {\it pure resource competition} scenario $g(x)=1$, in which $x$ and $N$ are coupled only through fluctuations; and
(ii) the {\it public good scenario} in which $g(x)=1+bx$, corresponding to a situation  where $S$ individuals are
``cooperators''~\cite{Roughgarden79,Melbinger2010,Cremer2011,Cremer09} producing a  public good (PG) that enhances 
the population growth rate through the benefit parameter $0<b= {\cal O}(1)$,
 here
assumed for simplicity to be independent of $s$.
In the PG scenario, $N$ and $x$ are explicitly coupled, since the changes in
the size of the population (\textit{ecological dynamics}) and those in its composition (\textit{evolutionary dynamics}) are interconnected.
This interplay establishes a form of ``eco-evolutionary dynamics''~\cite{Pelletier09,Harrington14}.
It is worth noting that, as customary in evolutionary game theory, we assume that mutation rates of the strains are negligible, 
and we thus characterize the population evolutionary dynamics in terms of the fixation properties~\cite{Nowak,Cremer09}.

In this context, the population size and composition  change according to the continuous-time birth-death 
process~\cite{Gardiner,vanKampen,Melbinger2010}
\begin{eqnarray}
\label{eq:rates}
N_{S/F}  \xrightarrow{T_{S/F}^{+}}  N_{S/F}+ 1 \quad \text{and} \quad
N_{S/F}  \xrightarrow{T_{S/F}^{-}}  N_{S/F}- 1,
 \end{eqnarray}
with transition rates
\begin{eqnarray}
\label{eq:ratesT}
T_{S/F}^{+}= g(x)\frac{f_{S/F}}{\bar{f}} N_{S/F} \quad \text{and} \quad 
T_{S/F}^{-}= \frac{N}{K(t)} N_{S/F}.
\end{eqnarray}

We model environmental randomness by letting the carrying capacity $K(t)$ switch randomly between 
 $K_+$ (abundant resources) and $K_-<K_+$ (scarce resources), see figure \ref{fig:Fig1}(a,b). We assume that
$K(t)$ switches at rate $\nu$, according to a  time-continuous symmetric dichotomous Markov noise (DMN)~\cite{HL06,Bena06,Kithara79}
$\xi(t)\in\{-1,+1\}$ (or {\it random telegraph noise}):
\begin{eqnarray}
\label{eq:xi}
\xi  \xrightarrow{\nu}  -\xi\,,
\end{eqnarray}
The stationary symmetric DMN has  zero-mean $\langle \xi(t)\rangle=0$ and autocorrelation $\langle \xi(t) \xi(t')\rangle={\rm exp}(-2\nu|t-t'|)$
($\langle \cdot\rangle$ denotes the ensemble average over the environmental noise)~\cite{HL06,Bena06}.
This is a {\it colored noise} with a finite correlation time $1/(2\nu)$~\cite{HL06,Bena06,Kithara79,Hanggi81,Broek84,Sancho85},
see section~\ref{SM1} in SM \cite{Supp}.
As a result, the fluctuating carrying capacity reads
\begin{eqnarray}
\label{eq:K(t)}
K(t)=\frac{1}{2}\left[(K_+ + K_-)+\xi(t) (K_+ - K_-)\right],
\end{eqnarray}
and endlessly switches between $K_{+}$ and $K_{-}$.

In what follows, we consider that the DMN is stationary: $\langle  \xi(t)\rangle=0$ for $t\geq 0$.
Hence, the initial carrying capacity is either $K_-$ or $K_+$ with probability $1/2$, and the 
average carrying capacity is constant: $\langle K(t)\rangle=\langle K\rangle=(K_+ + K_-)/2$.

The DMN models suddenly changing conditions, reflecting several
situations in bacterial life, such as cells living at either side of a physical
phase transition~\cite{Morley83}, or in the ever-changing conditions of
a host digestive tract.
In the laboratory, bacteria can be subjected to complex gut-like environment
\cite{Cremer17} or simplified stressful conditions, typically through
variable exposure to antibiotics \cite{Lindsey13,Lambert15,Lechner16}.
Furthermore, with modern bioengineering techniques it is possible to perform controled microbial experiments in 
settings allowing for sensible comparisons with theoretical models sharing some of the features 
considered here (switching environment, 
time-varying population size, public good production)~\cite{Acar08,Kussell05,Leibler09,Wienand15}.
As discussed in Section IV.B, the setting where colonies of bacteria are grown in arrays of wells or test 
tubes~\cite{Leibler09,Wienand15}, modeled 
as a metapopulation of communities, is particularly relevant for our purposes.

In this model, the population evolves according to the multivariate stochastic process defined by
equation~(\ref{eq:rates})-(\ref{eq:K(t)}), which obeys the master equation
\begin{eqnarray}
\label{eq:ME}
\frac{d P({\vec N},\xi,t)}{dt}&=&(\mathbb{E}^{-}_{S}-1)[T_{S}^{+} P({\vec N},\xi,t)] 
\nonumber\\ &+& 
(\mathbb{E}^{-}_{F}-1)[T_{F}^{+} P({\vec N},\xi,t)]\nonumber\\
&+&
(\mathbb{E}^{+}_{S}-1)[T_{S}^{-} P({\vec N},\xi,t)] ]\nonumber\\
&+&
(\mathbb{E}^{+}_{F}-1)[T_{F}^{-} P({\vec N},\xi,t)]\nonumber\\&+& 
\nu [P({\vec N},-\xi,t)-P({\vec N},\xi,t)],
\end{eqnarray}
where ${\vec N}=(N_S,N_F)$, $\mathbb{E}^{\pm}_{S/F}$ are shift operators such that
$\mathbb{E}^{\pm}_{S}G(N_S,N_F,\xi,t)= G(N_S\pm 1,N_F,\xi,t)$ for any $G(N_S,N_F,\xi,t)$,  and similarly for  $\mathbb{E}^{\pm}_{F}$.

Equation (\ref{eq:ME}) fully describes the stochastic eco-evolutionary 
dynamics of the population, and can be simulated exactly (see Sec.~\ref{AppendixB} in SM \cite{Supp}).

Importantly, here demographic fluctuations are
{\it coupled} to the {\it colored non-Gaussian}  environmental noise~\cite{KEM1,Supp} and encoded in the master equation (\ref{eq:ME}).
This contrasts with the discrete-time population dynamics of, for example Ref.~\cite{Loreau08},
where external and internal noises are  {\it independent and Gaussian}.
Simulation results, see figure~\ref{fig:Fig1}(c) and Ref.~\cite{videos}
(in which $N(0)=\langle K \rangle$, as in all our simulations), reveal that generally $N(t)$ evolves 
much faster than the population
composition.
We consider $K_+>K_-\gg 1$ to ensure that, after a transient, $N(t)$ is
at quasi-stationarity where it is characterized by
its quasi-stationary distribution ($N$-QSD). The population eventually collapses after a time that diverges with the system size~\cite{Spalding17,meta},
a phenomenon that can be disregarded for our purposes.
Below we study the eco-evolutionary dynamics in terms of the random variables $N$ and $x$, focusing on the
fixation properties of the population and its quasi-stationary distribution.

It is useful to start our analysis by considering the mean-field approximation which ignores {\it all noise} (say $K=\langle K \rangle$).
In this case, the population size $N$ and composition $x$ evolve deterministically according 
to~\cite{Melbinger2010,Cremer2011,KEM1,SM}
\begin{eqnarray}
\label{eq:N-deterministic}
\dot{N}&=& \sum_{\alpha=S,F}T_{\alpha}^+ -T_{\alpha}^- =N\left(g(x)-\frac{N}{K}\right)\,,\\
\label{eq:x-deterministic}
\dot{x}&=&  \frac{T_{S}^+ -T_{S}^-}{N} -x\frac{\dot{N}}{N}=-sg(x)\frac{x(1-x)}{1-sx}\,,
\end{eqnarray}
where the dot signifies the time derivative. Equation~(\ref{eq:x-deterministic}),  reminiscent of
 a replicator equation~\cite{Nowak},
predicts that $x$ relaxes on a timescale $t\sim 1/s\gg 1$ and eventually vanishes
while, according to equation~({\ref{eq:N-deterministic}}), $N(t)$ equilibrates to $N(t)= {\cal O}(K)$ in a time $t= {\cal O}(1)$.

\section{Piecewise-deterministic Markov process, Moran model \& Fixation probabilities}
\label{sec:PDMP-Moran-fixation}
In this section, we review  the  effects of environmental and demographic noise separately, and compound them to find 
the fixation probabilities that characterize the population composition.
Here, these  results provide the necessary background for the discussion  in Sections \ref{sec:public-good} and
\ref{INEN} of  our main novel findings.
\subsection{Environmental noise \&  Piecewise-deterministic Markov process}
\label{sub:PDMP}
If the population is only subject to external noise (EN), it follows the {\it bivariate} piecewise-deterministic Markov process
(PDMP)~\cite{Davis84}, defined by (\ref{eq:x-deterministic}) and
\begin{eqnarray}
\dot{N}&=& N\left\{g(x)-\frac{N}{{\cal K}}+
\xi N\left(\frac{1}{{\cal K}}-\frac{1}{K_+}\right)
\right\}\,,
\label{eq:N-stochastic}
\end{eqnarray}
where ${\cal K}=2K_+K_-/(K_+ + K_-)$ is the harmonic mean of $K_+$ and $K_-$ \cite{KEM1}. 
Equation (\ref{eq:N-stochastic}) is a stochastic differential equation with  multiplicative DMN $\xi$ of
amplitude $N^2(K_+ -K_-)/(2K_+K_-)$ \cite{Supp}; it reduces to the deterministic limit (\ref{eq:N-deterministic}) 
when the EN is removed (i.e. $K_+ =K_-$).

Although the process is only subject to EN, the global growth rate $g(x)$ couples the evolutionary and ecological dynamics.
To simplify the analysis, we introduce an effective parameter $q\geq 0$ (see Section \ref{sub:fixations-b}) and assume a constant $g\equiv 1+q$ \cite{KEM1}, 
obtaining the single-variate effective process
\begin{eqnarray}
\label{eq:sde}
&&\dot{N}= {\cal F}(N,\xi)=
\begin{cases}
{\cal F}_+(N)  &\mbox{if } \xi=1 \\
{\cal F}_-(N) &\mbox{if } \xi=-1, 
\end{cases}\\
&&\quad \text{with} \quad
{\cal F}_{\pm}(N) \equiv N\left[1+q-\frac{N}{K_{\pm}}\right],
\label{eq:fpm}
\end{eqnarray}
describing the evolution of a population of size $N(t)$ subject only to EN.
According to (\ref{eq:sde}) and (\ref{eq:fpm}), each environmental state $\xi$ has a fixed point
\begin{eqnarray}
  \label{eq:Npm}
 	N^*(\xi)=
	 \begin{cases}
 N^*_+=(1+q)K_+  &\mbox{if } \xi=1 \\
N^*_-=(1+q)K_- &\mbox{if } \xi=-1, 
 \end{cases}
\end{eqnarray}
After $t={\cal O}(1)$, the PDMP is at stationarity, characterized by a stationary probability density function (PDF)
$p^*_{\nu,q}(N,\xi)$ (derived in Section~\ref{AppendixC}~\cite{Supp}).
Central for our purposes are the features of the marginal stationary PDF $p^*_{\nu,q}(N)=p_{\nu,q}^*(N,\xi)+p_{\nu,q}^*(N,-\xi)$, giving the probability density of $N$ regardless of the environmental state $\xi$:
\begin{eqnarray}
\label{eq:pstarqmar}
p_{\nu,q}^*(N)
=\frac{{\cal Z}_{\nu,q}}{N^2}~\left[\frac{(N_+^*-N)(N-N_-^*)}{N^2}\right]^{\frac{\nu}{1+q}-1}\,,
\end{eqnarray}
with normalization constant ${\cal Z}_{\nu,q}$.
Depending on the sign of the exponent, the distribution may be unimodal or bimodal \cite{KEM1}, but has always support $[N_{-}^*,N_{+}^*]$, on which ${\cal F}_+\geq 0$ and ${\cal F}_-\leq 0$.

\subsection{Internal noise \& Fitness-dependent Moran process}
Internal noise stems from the inherent stochasticity of individual birth and death events in the population; it ultimately causes fixation 
(one strain taking over the whole population), and hence determines the long-term population composition.
When internal and ecological dynamics are coupled, which strain fixates has consequences on the population size, 
making the fixation phenomenon particularly important.

If internal noise is the only source of randomness (constant $K$), we can study its effects using the fitness-dependent Moran model \cite{Moran,Kimura,Blythe07,Cremer09,Antal06}, with constant
size $N\equiv K$ \cite{Otto97}.
To keep the population size constant, at each birth corresponds a death.
Therefore, $x$ increases by $1/N$ if an $S$ individual is born and an $F$ dies ($SF \to SS$  at rate $\widetilde{T}^{+}_S=T_S^{+}T_F^{-}/N$), and decreases by $1/N$ if an $F$ individual is born, replacing a dead $S$ ($SF \to FF$ at rate $\widetilde{T}^{-}_S= T_S^{-}T_F^{+}/N$), with
\begin{eqnarray*}
\hspace{-2mm}
 \label{tildeT}
 \widetilde{T}^{+}_S 
 =\frac{1-s}{1-sx}g(x)(1-x)x N, \;\;
 \widetilde{T}^{-}_S 
 =\frac{1}{1-sx}g(x)(1-x)x N\,.
 \end{eqnarray*}
The corresponding  mean-field equation is again  (\ref{eq:x-deterministic}).
For an initial fraction $x_0$ of $S$ individuals, in the framework of the Fokker-Planck equation, the fixation probability of $S$ is 
\cite{Gardiner,Kimura,Blythe07,Cremer09} (see also Section \ref{AppendixD}.A~\cite{Supp})
\begin{eqnarray}
\phi(x_0)|_{N}&=&\frac{e^{-N s(1-x_0)}-e^{-N s}}{1-e^{-N s}}\,.
\label{eq:phiNc}
\end{eqnarray} 
The fixation probability of $S$ thus becomes exponentially smaller the larger the population's (constant)
size or selection intensity $s$ are; and, notably, is independent of $g(x)$.
In the following we assume $x_0=1/2$ and drop the initial condition for notational simplicity: $\phi|_{N}\equiv \phi(x_0)|_{N}$ and
$\phi\equiv\phi(x_0)$.
Clearly, the fixation probability of $F$ is $\widetilde{\phi}|_{N}=1-\phi|_{N}$. 
In Section \ref{AppendixD}.A~\cite{Supp},
we also outline the main properties of the mean fixation times of the fitness-dependent Moran model.
The most relevant for our purposes is the fact that, in both  cases $b=0$ and $b>0$, the unconditional and conditional
mean fixation times scale as ${\cal O}(1/s)$ to leading order when $s\ll 1$ and $Ns\gg 1$.

\subsection{Fixation under switching carrying capacity}
\label{sec:fixations}
The strain $S$ unavoidably goes extinct in the deterministic limit, see equation (\ref{eq:x-deterministic}), and has an 
exponentially vanishing survival probability when $K$ is constant, see equation (\ref{eq:phiNc}). 
However, when the carrying capacity switches, the population undergoes ``bottlenecks'' that can 
enhance this probability \cite{KEM1} and alter the long-term average population size.

\subsubsection{Fixation probabilities in the pure competition scenario ($b=0$)}
\begin{figure}[htb]
\includegraphics{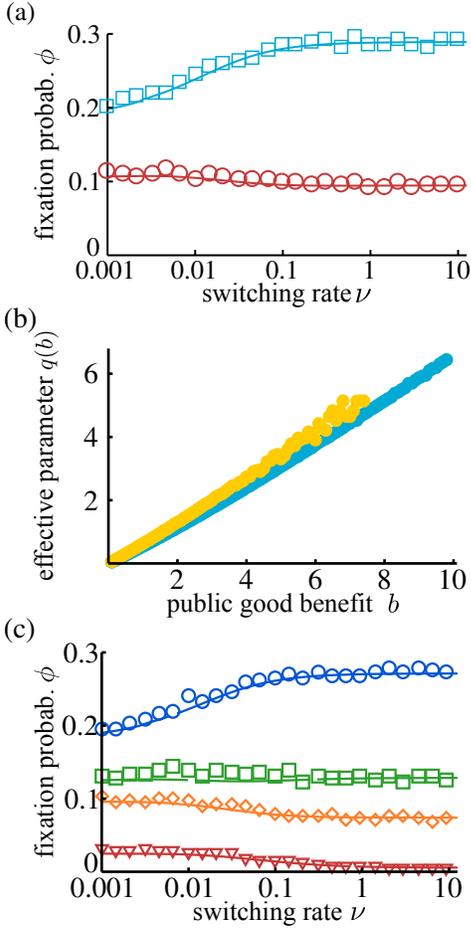}
\caption{
(a) $\phi$ vs. $\nu$ in the case $b=0$: for $s=0.02$ ($\Box$, cyan) and $s=0.05$ ($\circ$, red).
(b) $q(b)$ vs. $b$ for $s=0.02$ (cyan) and $s=0.05$ (yellow), see text. 
(c) $\phi$ vs. $\nu$ in the case $b>0$: for $(s,b)=(0.02,0.2)$ (blue, $\circ$), $(0.02,2)$ (green, $\square$), $(0.05,0.2)$ (orange, $\diamond$), $(0.05,2)$ (red, $\nabla$).
Symbols are $\phi$ from simulations ($10^4$ runs) and solid lines show $\phi_q$  from  equation~(\ref{eq:phi-nu_q}). 
In all panels, other parameters are $(K_{+},K_-,x_0)=(450,50,0.5)$.
} 
\label{fig:fixations}
\end{figure}

When $b=0$, both species compete for the same finite resources, with a slight selective advantage to $F$.
Therefore, $N$ and $x$ are solely coupled by demographic fluctuations.
After a time $t={\cal O}(1)$, $N$ attains quasi-stationarity 
where it is distributed according to its $N$-QSD~\cite{videos}, that is well 
described by the PDF of equation (\ref{eq:pstarqmar}) with $q=0$.
On the other hand, $x$ relaxes on a much slower timescale $t \sim 1/s\gg 1$
and we showed that the mean fixation time scales as ${\cal O}(1/s)$ to 
leading order when $s\ll 1$ and $\langle K \rangle s \gg 1$~\cite{KEM1,SM,Supp}. As a consequence, as shown in Section~\ref{AppendixD}.B~\cite{Supp},
the population experiences, on average, ${\cal O}(\nu/s)$ environmental switches prior
to fixation (see figures \ref{fig:avs}~\cite{Supp} and
\ref{fig:Fig1}(c)). When $0<s\ll 1$
 and $K_-\gg 1$, we can thus exploit this timescale separation and compute the $S$ fixation probability
$\phi$ by averaging $\phi|_N$ over the PDF $p_{\nu/s}^*\equiv p_{\nu/s,0}^*$, with the 
rescaled switching rate $\nu \to \nu/s$ \cite{KEM1}:
\begin{eqnarray}
\label{eq:phi-nu}
\phi&\simeq& \int_{K_-}^{K_+} \phi|_N~p_{\nu/s}^*(N)~dN.
\end{eqnarray}
The PDF $p_{\nu/s}^*$ is sharply peaked at $N\simeq {\cal K}$ when $\nu \gg s$, whereas it
has two sharp peaks at $N\simeq  K_{\pm}$ when $\nu\ll s$. 
Equation (\ref{eq:phi-nu}) captures the limiting behavior $ \phi \xrightarrow{\nu \to \infty}
\phi|_{\cal K}$ when  $\nu \gg s$ (many switches prior 
to fixation), resulting from the self-average of the EN (since $\xi(t) \xrightarrow{\nu \to \infty} 
\langle \xi(t)\rangle=0$), as well as $\phi \xrightarrow{\nu \to 0}
(\phi|_{K_-}+\phi|_{K_+})/2$ in the regime of rare switching ($\nu \ll s$), when the  environment almost never 
changes prior to fixation \cite{KEM1}.
As shown in figure \ref{fig:fixations} and detailed in Section \ref{AppendixAssess}~\cite{Supp}, 
equation (\ref{eq:phi-nu}) 
reproduces the simulation results 
for the fixation probability of $S$ within a few percent over a broad range of $\nu$ values.
While $S$ remains less likely to fixate than $F$, its fixation probability is much higher than in a constant environment 
($\phi\gg \phi|_{\langle K \rangle}$): environmental variability considerably offsets the evolutionary bias favoring $F$.

\subsubsection{Fixation in the public good scenario, $b>0$}
\label{sub:fixations-b}
In the public good scenario, $g(x)=1+bx$ with $0<b= {\cal O}(1)$, $S$ individuals act as public good producers (cooperators).
The higher $x$, in fact, the higher the reproduction rate of both strains, see equations (\ref{eq:ratesT}).
However, since $S$ bears alone  the metabolic cost of cooperation (PG production), it grows slower than $F$ and, deterministically, $x$  decreases.

When $b>0$, $N$ and $x$ are explicitly coupled, and they do not evolve on separate  timescales: $N$ is a fast variable, enslaved to the 
slow-varying $x$ \cite{videos}.
To determine the fixation probability, in Ref. \cite{KEM1} we devised an effective approach, based on suitably choosing the parameter 
$q$ ($0\leq q\leq b$) and setting $g(x)\equiv 1+q$ in equation (\ref{eq:N-stochastic}).
This decouples $N$ and $x$ in an effective population whose size distribution, at quasi-stationarity and for any $\nu$,
 is well described by the PDF (\ref{eq:pstarqmar}).
 Within this effective theory approach, the fixation probability of $S$ is thus determined similarly to the case $b=0$:
\begin{eqnarray}
\label{eq:phi-nu_q}
\phi_{q}&=& \int_{N_{-}^*}^{N_{+}^*}\!\!  \phi|_N~ 
p_{\nu/s,q}^*(N)~dN\,.
\end{eqnarray}
As above, As above, this expression simplifies in the limiting regimes of frequent/rare switching:
$\phi_{q}\simeq \phi_{q}^{(\infty)}\equiv \phi|_{(1+q){\cal K}}$
when $\nu\gg s$, and 
$\phi_{q}\simeq \phi_{q}^{(0)}\equiv (\phi|_{N_{-}^*}+ \phi|_{N_{+}^*})/2$ when $\nu\ll s$.
We determined the effective parameter $q=q(b)$ for given $(K_\pm, s, b)$ by matching the prediction of $\phi_{q}^{(\infty)}$ with the results
of simulations (see \cite{KEM1} and SM~\cite{Supp}).
Figure \ref{fig:fixations}(b) shows that $q(b)$ increases almost linearly with $b$, and depends weakly on $s$.
Clearly, $q(0)=0$ when $b=0$, and equation (\ref{eq:phi-nu_q}) thus reverts to (\ref{eq:phi-nu}).

Figure \ref{fig:fixations}(c) shows that the effective approach captures the effects of the coupling between $N$ and $x$ for several 
choices of $b$ and $s$, over a broad range of $\nu$.
As detailed in \ref{AppendixAssess}, the predictions of equation (\ref{eq:phi-nu_q}) agree within a few percent with simulation results 
when $s\ll1$, while the accuracy deteriorates as $s$ and $b$ increase, therefore lowering $\phi$.
In fact, increasing $b$ yields higher $q(b)$, which results in effectively increasing the carrying capacity $K_{\pm} \to (1+q(b))K_{\pm}$.
In the $\nu \to \infty, 0$ limits, this is equivalent to rescaling the selection intensity as
$s \to (1+q(b))s$, as inferred from $\phi_{q}^{(\infty, 0)}$ and equation (\ref{eq:phiNc}).
Therefore $\phi$ decays (approximately) exponentially with $b$, as shown by figure~\ref{fig:pdmp-predictions}(a). 

\section{Correlations \& cooperation in the eco-evolutionary game}
\label{sec:public-good}

After a time $t\gg 1/s$, fixation has very likely occurred and the population composition is fixed and 
consists of only $F$ or $S$ individuals. In this quasi-stationary regime, the population  size $N(t)$ 
however keeps fluctuating, driven by the randomly switching carrying capacity $K(t)$.

When the slow strain $S$ produces a public good (PG), the long-time eco-evolutionary dynamics is 
characterized by correlations between the population size and its composition.
In this section, we analyze these long-term effects by characterizing the correlations first,
and then by analyzing the ensuing
``eco-evolutionary game''.

\begin{figure}
\begin{center}
\includegraphics{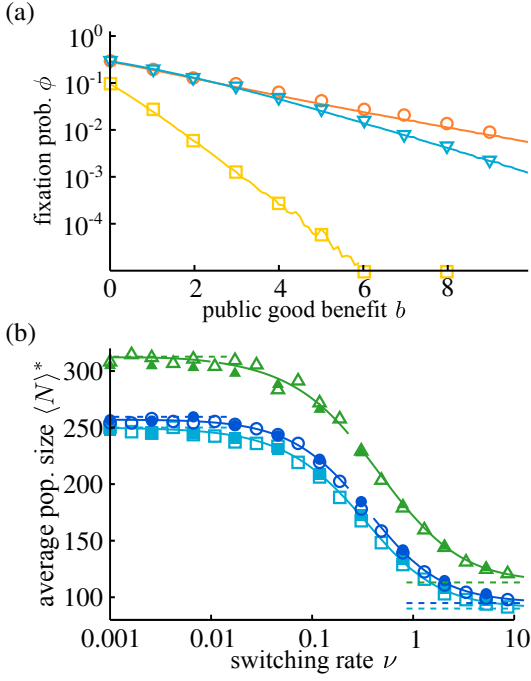}
\end{center}
\caption{
(a) $\phi$ vs $b$ in lin-log scale for $s=0.02$, $\nu=0.1$ (orange, $\circ$) and $\nu=10$ (cyan, $\nabla$); $s=0.05$, $\nu=10$ (yellow, $\Box$).
Lines are from (\ref{eq:phi-nu_q}) and markers are from simulations.
(b) $\langle N \rangle^*_{\nu,b}$ vs. $\nu$ for $b=0$ (cyan, squares), $b=0.2$ (blue, circles) and $b=2$ (green, triangles) and $s=0.02$. 
Solid lines are from (\ref{eq:Nstarq}); empty symbols are from simulations; filled symbols are from  (\ref{eq:av_LNA})
 within the linear noise approximation. Dashed lines indicate the predictions of (\ref{eq:Nstarq}) in the 
 regimes $\nu \to \infty, 0$, 
 see \ref{AppendixPDMP}.A~\cite{Supp}. Parameters are $(K_+,K_-,x_0)=(450,50,0.5)$.}
\label{fig:pdmp-predictions}
\end{figure}

To this end, it is useful to consider the average population size $\langle N \rangle_{\nu,b}^*$ for given $\nu$ and $b$, after a time 
$t\gg 1/s$, when the population is at \emph{quasi-stationarity} and consists of only $S$ or $F$ individuals, see Section \ref{AppendixD}.B~\cite{Supp}.
Within the PDMP approximation---that is, approximating the evolution of $N$
by the PDMP (\ref{eq:sde}), see Section \ref{AppendixPDMP}.A~\cite{Supp}---we can compute the quasi-stationary average of $N$
using $p^*_{\nu,q}$ given by eq.~(\ref{eq:pstarqmar}) 
 (see also Sec.~\ref{sec:LNA}):
\begin{eqnarray}
\label{eq:Nstarq}
\hspace{-6.5mm}
\langle N  \rangle_{\nu,b}^*= (1+b)
\phi_b\langle N   \rangle_{\frac{\nu}{1+b},0}^* + \widetilde{\phi}_b\langle N   \rangle_{\nu,0}^*>
\langle N   \rangle_{\nu,0}^*,
\end{eqnarray}
where $\langle N   \rangle_{\nu,0}^*$ is the population long-time average in the absence of PG production, $\phi_b$  
denotes the fixation probability of $S$ for a public good parameter $b$, and $\widetilde{\phi}_b=1-\phi_b$.
Through equation (\ref{eq:Nstarq}), the PDMP approximation thus predicts that the long-term population 
size increases with $b$, see figure \ref{fig:pdmp-predictions}(b). Furthermore, while the fixation probabilities 
$\phi_b$ and  $\widetilde{\phi}_b$ can increase or decrease with $\nu$, the PDMP approximation predicts that the 
average population size at stationarity monotonically decreases with $\nu$ (see (S20)-(S22) in \ref{AppendixPDMP}.A~\cite{Supp}).
Simulation results shown in figure \ref{fig:pdmp-predictions}(b) confirm that 
 $\langle N  \rangle_{\nu,b}^*$  increases with $b$, and decreases with $\nu$ (keeping other parameters constant).

\subsection{Correlations between ecological \& evolutionary dynamics}
Equation (\ref{eq:Nstarq}) also highlights how fixation probabilities affect the long-term average population size.
When $b>0$, there are nontrivial correlations between population size and composition, and how 
$N(t)$ and $x(t)$ are correlated is  
of direct biological relevance, see {\it e.g.}~\cite{Leibler09,Melbinger2015a}.
Prior to fixation, these correlations are accounted by the effective parameter $q(b)$ (see section \ref{sub:fixations-b}).
Here, we investigate their effect after fixation using the rescaled connected correlation function
\begin{eqnarray}
\label{eq:conn}
{\cal C}_{\nu,b}(t)=\frac{\left\langle \left(  N(t)- \langle N (t)\rangle \right)\left( x(t)- \langle x(t) \rangle\right)
\right \rangle }{\langle N (t)\rangle \langle x(t) \rangle}\,,
\end{eqnarray}
where $\left\langle \cdot \right\rangle$ denotes the ensemble average.
When $\langle N (t) x(t) \rangle=\langle N (t)\rangle \langle x (t) \rangle$, i.e.
in absence of correlations, ${\cal C}_{\nu,b}(t)$ vanishes.
At quasi-stationary, $t\gg 1/s$, we have $\langle N (t) x(t)\rangle \to \langle N  x\rangle_{\nu,b}^*$,
$\langle N (t)\rangle \to \langle N\rangle_{\nu,b}^*$, $x \to 1$ or $0$ with respective probability 
$\phi_b$ and $\widetilde{\phi}_b$,
$\langle x(t)\rangle \to \phi_{b}$ and ${\cal C}_{\nu,b}(t) \to {\cal C}_{\nu,b}^*$.
Within the PDMP approximation, using eq.~(\ref{eq:Nstarq}) and $\phi_b\simeq \phi_{q}$, 
equation (\ref{eq:conn}) becomes ($t\gg 1/s$)
\begin{eqnarray}
\hspace{-3.5mm}
 \label{eq:Corr}
 {\cal C}_{\nu,b}^*=\frac{\langle N x \rangle^*_{\nu,b}}
 {\langle N \rangle^*_{\nu,b} \phi_b}-1
 \simeq \frac{\widetilde{\phi}_{q} \left[(1+b)\langle N\rangle^*_{\frac{\nu}{1+b},0} - 
 \langle N\rangle^*_{\nu,0}\right]}{(1+b)\phi_{q}\langle N\rangle^*_{\frac{\nu}{1+b},0}+
 \widetilde{\phi}_{q}\langle N\rangle^*_{\nu,0}}.\,
\end{eqnarray}
Since $\langle N\rangle^*_{\nu,0}$ is decreasing in $\nu$ (see figure \ref{fig:pdmp-predictions}(a)), this 
long-term correlation is always positive for $b\geq0$, and vanishes only for $b=0$.

\begin{figure}
\begin{center}
\includegraphics{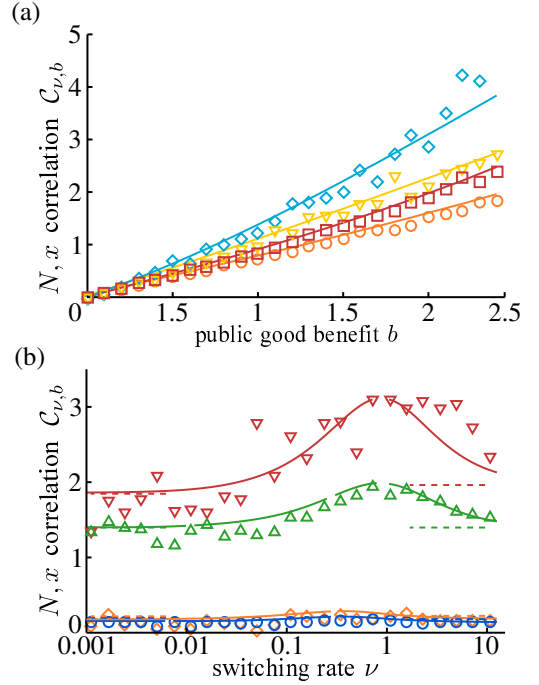}
\caption{
(a) ${\cal C}_{\nu,b}^*$ vs $b$ 
for $s=0.05$ and  $\nu\simeq1$ (cyan, $\diamond$), $\nu=0.1$ (yellow, $\nabla$); $s=0.02$ and $\nu=1$ (red, $\square$), and $\nu=0.1$ (orange, $\circ$).
(b) ${\cal C}_{\nu,b}^*$ vs $\nu$ for $b=2$ and $s=0.05$ (red, $\nabla$), $s=0.02$ (green, $\triangle$); $b=0.2$ and $s=0.05$ (orange, $\diamond$), $s=0.02$ (blue, $\circ$).
In all panels, the parameters are $(K_+,K_-,x_0)=(450,50,0.5)$.
Symbols are results from simulations and solid lines are from equation~(\ref{eq:Corr}); dashed lines in panel (b) denote the analytical predictions of ${\cal C}_{\nu,b}$ in the limits $\nu\ll s$ and $\nu\gg 1$, see text.
}
\label{fig:corr}
\end{center}
\end{figure}

As shown in figure~\ref{fig:corr}, ${\cal C}_{\nu,b}^*$ grows approximately linearly with $b$ and is non-monotonic in $\nu$ with a maximum for $\nu={\cal O}(1)$; all features that equation (\ref{eq:Corr}) captures well.
The $\nu$-dependence of ${\cal C}_{\nu,b}^*$ stems from the fact that $\phi_b$ increases or decreases with $\nu$, depending on the value of $s$, see figure \ref{fig:fixations}(c) \cite{KEM1}.
In the limiting regimes $\nu \to \infty,0$, equation (\ref{eq:Corr}) simplifies and yields ${\cal C}_{\nu,b}^*\simeq b[1-(1+b)\phi_{q(b)}^{(\infty,0)}]$ \cite{Supp}.
Therefore, in these the limiting regimes ${\cal C}_{\nu,b}^*$ increases in $s$, and scales as ${\cal O}(b)$, as shown 
by figure~\ref{fig:corr}, yielding $\langle N x\rangle^*=
(1+{\cal O}(b))\langle N\rangle^*_{\nu,b}\phi_{q(b)}^{(\infty,0)}$.

These results show that, when species $S$ provides a PG, there are nontrivial long-term correlations between \emph{ecological and evolutionary variables}: the population size is shaped by its composition.
The correlations between $N$ and $x$ are maximal in the intermediate switching regime where $\nu={\cal O}(1)$ is comparable to the growth rate of $N$, and are weaker in the limiting switching regimes, on which we devised the effective theory of section \ref{sub:fixations-b}.

\subsection{When is cooperation beneficial? In which conditions is it best to cooperate?}
Producing the public good (PG) slows the growth of the $S$ strain, see equation (\ref{eq:x-deterministic}) with $g(x)>0$, and thus reduces exponentially
the fixation probability of $S$, as shown in figure \ref{fig:pdmp-predictions}(a).
On the other hand, the PG leads to higher average population sizes (see equation (\ref{eq:Nstarq}) and 
figure \ref{fig:pdmp-predictions}(b)) and therefore provides a long-term benefit to the whole population.
At the population level, a ``social dilemma'' \cite{Nowak,Cremer09} of sorts arises after fixation of either $F$ or $S$:
Cooperators pay a cost through their reduced fixation probability, 
while they provide a benefit, through the PG, by increasing the expected long-term number of individuals of both strains.
We analyze the trade-off between benefit and cost of cooperation
by introducing the notion of ``eco-evolutionary game'' in the context of a metapopulation of non-interacting 
communities:
Each system realization (simulation run) corresponds to a community of time-fluctuating size,
and the collection of the system's realizations constitutes the metapopulation 
\cite{Leibler09,Melbinger2010,Cremer2011,Wienand18}, that is an ensemble of non-interacting communities.
After fixation, each community consists of only $S$ or $F$ individuals.
It is worth emphasizing that the social dilemma arising in the eco-evolutionary game
differs from traditional games in a finite population of constant size \cite{Nowak,Cremer09}:
Although $F$ is always more likely to fixate than $S$ (when $x_0$ is not too close to $1$, as in classical evolutionary games),
communities consisting only of individuals of strain $S$ can be of significantly larger size 
than those containing only $F$'s thanks to their production of PG (allowing them to possibly attain
the maximum carrying capacity $(1+b)K_+$).
In this eco-evolutionary game in a population of 
\emph{time-varying size}, we thus propose to measure the evolutionary success of a strain in terms of the 
population size averaged after fixation over the ensemble of non-interacting communities, see also 
 Section \ref{AppendixPDMP}.B of the SM~\cite{Supp}: The expected payoff of the game is hence the relative long-term average
number of individuals of each strain, see below.
Interestingly, this formulation of the eco-evolutionary game is of potential direct 
 relevance to microbial experiments in which colonies of bacteria, some of which can produce a public good,
are grown and compete in ``a metapopulation of test tubes'', see {\it e.g.}~\cite{Leibler09,Melbinger2015a,Wienand15,Wienand18}.

Below, we use the PDMP approximation and  simulations to investigate the relative abundance of each species at
quasi-stationarity (see also Section \ref{AppendixPDMP}.A~\cite{Supp}).

\begin{figure}
\begin{center}
\includegraphics{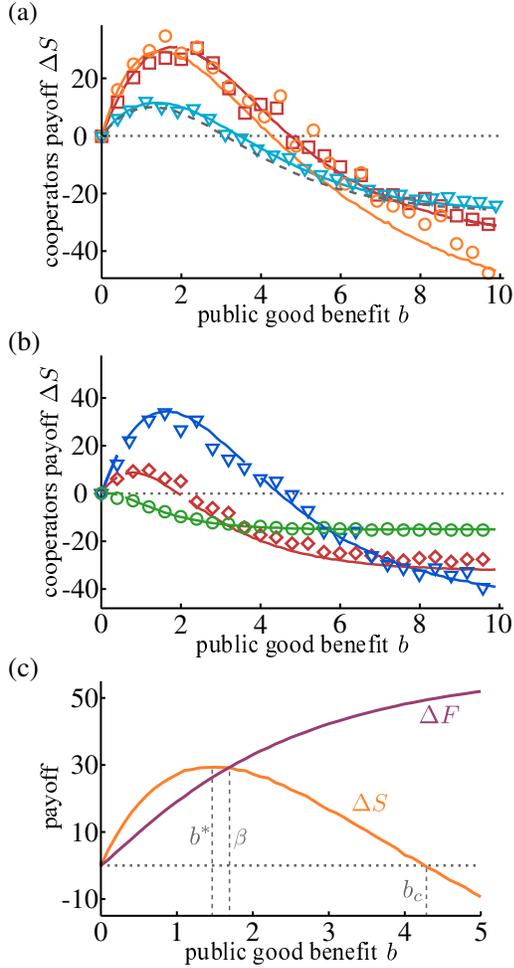}
\caption{
(a)
$\Delta S_{\nu, b}$ vs. $b$ for $s=0.02$ and switching rates $\nu=10$ (cyan, $\nabla$), $\nu=1$ (red, $\square$), 
$\nu=0.1$ (orange, $\circ$).
Predictions from equation (\ref{eq:NbN0}) (solid) are compared to simulation results (symbols).
The grey dashed line corresponds to the predictions of $\Delta S_{\infty, b}$, see text.
We find $\Delta S_{\nu,b} > 0$ when $0<b<b_c(\nu,s)$ with an optimal payoff for $S$ when $b=b^*(\nu,s)$, e.g. $(b_c,b^*)\approx (4.9, 2.1)$ at $\nu=1$.
(b) $\Delta S_{b,\nu}$ vs. $b$  with $\nu\simeq0.44$, for  $s=0.02$ (blue, $\nabla$), $s=0.03$ (red, $\diamond$), and $s=0.05$ (green, $\circ$).
Solid lines are from equation~(\ref{eq:NbN0}) and symbols are simulation results (see SM \cite{Supp}).
(c) Expected payoffs $\Delta S_{\nu, b}$ and $\Delta F_{\nu, b}$ vs. $b$ for $s=0.02$ obtained from equation (\ref{eq:NbN0}).
Dashed lines show the values of $b^*$, $\beta$ and $b_c$.
In all panels, the parameters are $(K_+,K_-,x_0)=(450,50,0.5)$.
}
 \label{fig:Delta}
 \end{center}
\end{figure}

The average number of $F$ individuals at quasi-stationarity, given a switching rate $\nu$ and PG parameter $b$ is
\begin{eqnarray}
 \label{NF}
 \nonumber
\langle N_F \rangle^*_{\nu,b}=\langle N| x=0 \rangle^*_{\nu,b}=(1-\phi_b)
 \langle N \rangle^*_{\nu,0},
\end{eqnarray}
i.e. the average population size conditioned to $F$ fixation. 
Similarly, the average number of cooperators $S$ at quasi-stationarity is 
\begin{eqnarray}
\nonumber
 \label{NS}
\langle N_S \rangle^*_{\nu,b}=\langle N| x=1 \rangle^*_{\nu,b}=(1+b)\phi_b \langle N \rangle^*_{\nu/(1+b),0}.
\end{eqnarray}
In the context of the above eco-evolutionary game, we propose to measure the expected payoff provided by the PG as \textit{the difference  between the expected number of individuals of a 
strain at quasi-stationarity when $b>0$ relative to the case $b=0$}. Hence, the expected payoff to $F$ is
\begin{eqnarray}
\label{eq:DF}
\Delta F_{\nu,b}\equiv \langle N_F \rangle^*_{\nu,b}-\langle N_F \rangle^*_{\nu,0}=
(\phi_0 - \phi_b)\langle N \rangle^*_{\nu,0}>0.
\end{eqnarray}
Since $\phi_0 >\phi_b$, see figure~\ref{fig:pdmp-predictions}(b), this quantity is positive and increases with $b$.
This means that, as in other social dilemmas, see, {\it e.g.}, Ref.~\cite{Nowak,Cremer09}, 
the benefit of ``freeriding'' increases
when the level  of cooperation, here given by $b$, is raised.
However, this does not rule out the possibility that, under certain circumstances, 
the PG production can be either beneficial or detrimental to $S$, and even permits $S$ 
to be better off than $F$. 
In fact, the (global) eco-evolutionary expected payoff for cooperators reads
 \begin{eqnarray}
  \label{eq:DS}
\hspace{-2mm}
  \Delta S_{\nu,b}&\equiv&\langle N_S \rangle^*_{\nu,b}-\langle N_S \rangle^*_{\nu,0}\nonumber\nonumber\\ &=&(1+b)
\phi_b\langle N  \rangle_{\frac{\nu}{1+b},0}^* -\phi_0\langle N  \rangle_{\nu,0}^*,
 \end{eqnarray}
and clearly varies nontrivially with  $\nu$ and $b$.
Unless $\Delta S_{\nu,b}>0$, the PG is actually detrimental for cooperators: the expected number of $S$ individuals is lower than 
it would be without PG.
In this context, the PG benefits cooperators only if the increase in the average population size offsets the decrease in fixation probability, i.e. if
\begin{eqnarray}
\label{eq:Delta}
\nonumber
\Delta S_{\nu,b}>0 \Leftrightarrow
(1+b)\frac{\langle N \rangle_{\frac{\nu}{1+b},0}^*}{\langle N \rangle_{\nu,0}^*}>\frac{\phi_0}{\phi_b}
\end{eqnarray}
In figure \ref{fig:Delta}, we show that $\Delta S_{\nu, b}$ is non-monotonic in $b$, generating a maximum at an \emph{optimal value} $b^*(\nu,s)$, which defines the 
conditions where PG production is the most rewarding for cooperators.
Moreover, we observe a definite \emph{critical threshold} $b_c(\nu,s)$, below which producing a PG benefits 
cooperators since $\Delta S_{\nu,b}>0$.

Using our effective theory, $\phi\simeq \phi_{q(b)}$, and the PDMP approximation, the expected payoff of $S$ (\ref{eq:DS}) reads
\begin{eqnarray}
\label{eq:NbN0}
\Delta S_{\nu, b}
&=&
(1+b)\phi_{q(b)}\int_{K_-}^{K_+}  N p_{\frac{\nu}{1+b}}^*(N)~dN\nonumber\\
&-&\phi_0\int_{K_-}^{K_+}  N p_{\nu}^*(N)~dN\,.
\end{eqnarray}
When $\nu\to \infty$, the DMN self-averages ($\xi \to \langle \xi \rangle=0$) and equation~(\ref{eq:DS}) is given by 
the  expected payoff of $S$ in a population of effective size 
$\langle N \rangle_{\infty,0}^*={\cal K}$, see equation~(S23)~\cite{Supp}, yielding
 $\Delta S_{\infty, b}=[(1+b)\phi_{q(b)}^{(\infty)} -\phi_0^{(\infty)}]{\cal K}$. Hence, 
 when the DMN self-averages, the  expected payoff of $S$ is positive
 if $\phi_{q(b)}^{(\infty)}>\phi_0^{(\infty)}/(1+b)$.

\begin{figure}
\begin{center}
\includegraphics{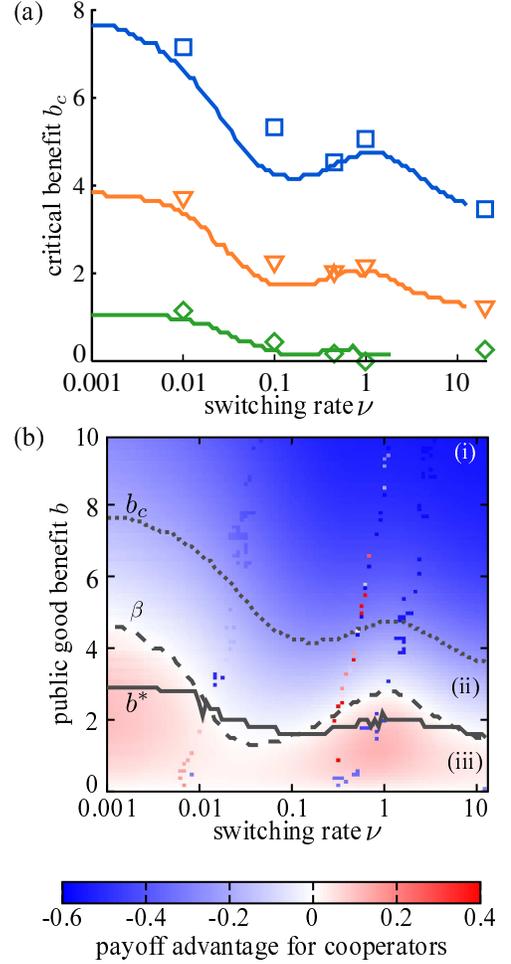}
 \end{center}
\caption{
(a) $b_c$ vs $\nu$.
Symbols are results from simulations and solid lines are from equation (\ref{eq:NbN0}) for $s=0.02$ (blue), $s=0.03$ (orange), and $s=0.05$ (green).
(b) Heatmap of $(\Delta S_{\nu,b}-\Delta F_{\nu,b})/\langle N\rangle_{\nu,0}^*$, from equation~(\ref{eq:NbN0}) for $s=0.02$.
The gray dotted line shows $b=b_c(\nu,s)$, the dashed line $b=\beta(\nu,s)$ and the solid line $b=b^*(\nu,s)$.
In the blue area (phases (i) and (ii)), $b>\beta$ and $F$ is better off than $S$ ($\Delta F_{\nu,b}> \Delta S_{\nu,b}$).
PG production is detrimental for $S$ in phase (i) where $b>b_c$ and $\Delta S_{\nu,b}<0$; beneficial for $S$ ($\Delta S_{\nu,b}>0$) in phase (ii) where $\beta<b<b_c$, but more beneficial for $F$ (higher expected payoff).
In the red/pink area of region (iii), $b<\beta$ and $S$ is better off than $F$ ($\Delta S_{\nu,b}> 
\Delta F_{\nu,b}$), see text.
Colored dots correspond to ``gaps'' in the numerical data (see \cite{Supp}). 
Parameters are $(K_+,K_-,x_0)=(450,50,0.5)$.
\label{fig:heatmap}
} 
\end{figure}

Results in figure \ref{fig:Delta} show that equation (\ref{eq:NbN0})  approximates well the simulation results over a 
broad range of parameters. The root and the maximum of equation (\ref{eq:NbN0}) provide
(approximate)
predictions for $b_c$ and $b^*$, see figures \ref{fig:heatmap} and S7(a)~\cite{Supp}. These 
figures reveal that $b_c$ and $b^*$ depend non-monotonically on $\nu$ and vary greatly with $s$, 
both behaviors well-captured by the theory.
Figures  \ref{fig:Delta} and S7(b)~\cite{Supp} also 
show that the maximal payoff for $S$ can be significantly higher than that of $F$, especially when the selection $s$ is low.

In order to discuss the properties of the eco-evolutionary game, it is  useful to determine the value $b=\beta(\nu,s)$ of equal expected payoff, i.e. such that
which $\Delta S_{\nu,\beta}= \Delta F_{\nu,\beta}$, see figure \ref{fig:Delta}(c).
From equations (\ref{eq:DF})-(\ref{eq:NbN0}), we find that  $\beta(\nu,s)$ is the solution of
\begin{eqnarray}
 \label{eq:beta}
\hspace{-3mm}
\frac{1}{1+\beta}\left(\frac{2\phi_0}{\phi_{q(\beta)}}-1\right)= \frac{\langle N \rangle_{\frac{\nu}{1+\beta},0}^*}{\langle N \rangle_{\nu,0}^*}
=\frac{\int_{K_-}^{K_+}N p^*_{\frac{\nu}{1+\beta}}~dN}{\int_{K_-}^{K_+}N p^*_{\nu}~dN}\,.
\end{eqnarray}
So $\beta$ is a nontrivial function of $\nu$ and $s$, see figure \ref{fig:heatmap}(b).

Given the parameters $b$ and $s$ of the eco-evolutionary game, we have studied
the values of the switching rate $\nu$ for which it is beneficial 
to cooperate by producing a public good, and determined three distinct phases 
represented in the diagram of figure \ref{fig:heatmap}(b):
\begin{enumerate}
\item[(i)] When $b>b_c$, the PG production is detrimental for $S$.
The cost of cooperation outweighs its benefits and the expected payoff for $S$ is negative ($\Delta S_{\nu,b}<0$). The PG thus benefits only $F$.
\item [(ii)] When $\beta<b<b_c$, the PG production benefits $S$, but benefits $F$ more ($0< \Delta S_{\nu,b} < \Delta F_{\nu,b}$).
\item [(iii)] When $0<b< \beta$, $S$ reaps a higher expected payoff than $F$ ($\Delta S_{\nu,b} > \Delta F_{\nu,b}>0$).
In this case, the benefit of the PG outweigh its cost, and its production is \emph{favored}.
\end{enumerate}

Within the above metapopulation interpretation of the eco-evolutionary game, species $F$ effectively exploits $S$ in 
phases (i) and (ii), but is at a disadvantage in phase (iii). 
Since the expected payoff to $S$ is positive in regions (ii) and (iii), we say that {\it cooperation
of a public good 
with benefit parameter $b$ is beneficial when $0<b<b_c(\nu,s)$, and advantageous for $0<b<\beta(\nu,s)$.}
Given a set of parameters $(b,\nu,s)$, PG production is the best strategy if two conditions are met: 
(a) the expected payoff of $S$ is higher than that of $F$, which is satisfied in phase (iii);
(b) $b$ yields the maximum possible payoff for $S$, i.e., $b=b^*$.
Hence,
{\it  in an environment switching at rate $\nu$ and under a selection intensity $s$, the best conditions to cooperate for the  public 
good production is when the PG benefit parameter satisfies $b=b^*(\nu,s)<\beta(\nu,s)$}, represented by the 
solid gray line in phase (iii) of figure~\ref{fig:heatmap}(b).
It is also worth noting that this discussion also holds when the time-varying population size
is not driven by the environmental noise: The limiting case 
$\nu \to \infty$, for which the DMN self-averages 
and the population reaches an effective size $N\to {\cal K}$,  corresponds to the right end of the diagram of figure~\ref{fig:heatmap}(b)
where $\nu\gg 1$. Remarkably, environmental stochasticity 
yields several additional regimes in which cooperating becomes beneficial.

In the above context of a well-mixed population whose size fluctuates in time, 
this eco-evolutionary game shows that there are conditions under which PG production is beneficial 
for cooperators, and may even be the optimal strategy. This does not imply that the social dilemma, 
which still holds in its classical form prior to fixation, is resolved in general. 
However, this demonstrates that under environmental variability there are conditions in which cooperation (PG production), 
albeit disadvantaged in the short term, can be more successful than freeriding in the long term.
In fact, although freeriders have a constant growth-rate advantage over cooperators and are always more 
likely to fixate (assuming $x_0=1/2$), here, the selective bias can be efficiently balanced by environmental variability, by allowing 
cooperators to be successful  in forming, in the long term, larger communities than freeriders~\cite{Supp}.
This can result in a greater increase of the long-term average number of cooperators than free-riders, and
exemplifies the potential role of a fluctuating environment on the emergence of cooperative behavior
in microbial communities.

\section{Linear-noise and PDMP approximations to the population QSD}
\label{INEN}
After $t\gg 1/s$, the population is likely to be at quasi-stationarity
with its composition fixed \cite{KEM1}.
Yet, the population size still fluctuates and $N(t)$ is distributed 
according to its quasi-stationary distribution. 
When $K_-\gg 1$, the population size is always large and, in the first instance, demographic fluctuations 
are negligible compared to environmental noise. In this case, eq.~(\ref{eq:sde}) characterizes reasonably well, 
albeit not fully, 
the long-term properties of $N(t)$.

\subsection{Linear-noise approximation about the PDMP predictions}
\label{sec:LNA}
Throughout this work (and in \cite{KEM1}), we have shown that the PDMP approximation 
$p^*_{{\rm PDMP},\nu,b}(N)=\phi p^*_{\nu,b}(N) + \widetilde{\phi}p^*_{\nu}(N)$ reproduces many characteristics of the quasi-stationary size distribution ($N$-QSD).
However, as $p^*_{\nu}$ and $p^*_{\nu,b}$ only account for the external noise (EN), they cannot reproduce the complete $N$-QSD, which is also subject to internal noise (IN).
Here, we use the linear noise approximation (LNA) about the PDMP predictions to account for the {\it joint effect} of the two noise sources, IN and EN, 
on the $N$-QSD.

The LNA is widely employed to quantify the effect of weak demographic fluctuations in the absence of external noise \cite{vanKampen,Gardiner}, and has recently been used to study the joint effect of decoupled internal and external noise \cite{Hufton16}.
Here, we show how to generalize the LNA to the case where the population size fluctuates and demographic fluctuations are coupled to the external noise.

For our analysis, we assume that $K_+\gtrsim K_-\gg 1$, so that $\langle K \rangle$ is large and of the same order as $K_{\pm}$ (see 
Section \ref{AppendixE}~\cite{Supp} for details).
It is convenient to work with the continuous random variable $n=N/\Omega$, where $\Omega=\langle K \rangle\gg 1$ is the system's ``large parameter''. 
The auxiliary Markovian process $\{n(t),\xi(t)\}$ that we consider for the LNA is defined by $n \stackrel{{\cal T}^+}
     {\longrightarrow} n+\Omega^{-1}, \quad n \stackrel{{\cal T}^-}
     {\longrightarrow} n-\Omega^{-1}$ and $\xi \stackrel{\nu}{\longrightarrow} -\xi$, 
where the transition rates ${\cal T}^{\pm}$ are given by equations~(S30) in the SM \cite{Supp}. 
We also introduce $\psi=\lim_{\Omega\to \infty} N/\Omega={\cal O}(1)$, which obeys the stochastic differential 
equation (S33) \cite{Supp} defining the corresponding PDMP, and
the random variable $\eta(t)$, capturing the fluctuations of $n$ about $\psi$, according to
\begin{eqnarray} 
\label{eq:n}
n(t)=\psi(t)+\frac{\eta(t)}{\sqrt{\Omega}},
\end{eqnarray}

We are interested in the (quasi-)stationary joint probability density $\pi^*_{\nu,q}(\eta, \psi, \xi)$ of the process $\{n(t),\xi(t)\}$.
This  probability density can be decomposed into $\pi_{\nu,q}^*(\eta, \psi, \xi)=\pi^*(\eta| \psi, \xi)\pi_{\nu,q}^*(\psi, \xi)$, 
where $\pi_{\nu,q}^*(\psi, \xi)=\Omega p^*_{\nu,q}(\Omega \psi,\xi)$ is the stationary joint PDF of the
PDMP $\{\psi(t)\}$ and is readily obtained from the PDF of the PDMP defined by equation (\ref{eq:sde}).
The stationary probability density $\pi^*(\eta| \psi, \xi)$ accounts for the demographic fluctuations about 
$\{\psi(t)\}$ in the environmental state $\xi$.
Following Ref.~\cite{Hufton16}, we assume that the demographic fluctuations are approximately the same in both
environmental states, i.e.  $\pi_{\nu,q}^*(\eta| \psi, \xi)\simeq \pi_{\nu,q}^*(\eta| \psi, -\xi)$, and 
simply denote $\pi_{\nu,q}^*(\eta| \psi)\equiv\pi_{\nu,q}^*(\eta|\psi,\pm\xi)$. 
This assumption is reasonable when $K_+$ and $K_-$ are of the same order, and yields
\begin{eqnarray}
\label{eq:approx}
\pi^*_{\nu,q}(\eta, \psi, \xi)\simeq \pi^*(\eta| \psi)\pi_{\nu,q}^*(\psi, \xi).
\end{eqnarray}
With this approximation, the  quasi-stationary marginal LNA  probability density of $\{n(t)\}$ is
\begin{eqnarray}
 \label{eq:PDF_LNA}
 \pi_{\nu,q}^*(n)&=&\sum_{\xi=\pm 1} \int \int d\psi d\eta~\pi^*(\eta| \psi) \nonumber\\
&\times&
 \pi_{\nu,q}^*(\psi, \xi)~\delta\left(n-\psi-\frac{\eta}
 {\sqrt{\Omega}}\right),
\end{eqnarray}
where $\pi^*(\eta| \psi)={\rm exp}\{-\eta^2/(2\psi)\}/\sqrt{2\pi\psi}$ (see SM \cite{Supp} for details), 
and the Dirac delta ensures that (\ref{eq:n}) is satisfied.
Calling $p_{{\rm LNA},\nu,0}^*(N)= \pi_{\nu,0}^*(n)/\Omega$ and $p_{{\rm LNA},\nu,b}^*(N)= \pi_{\nu,b}^*(n)/\Omega$,  explicitly given 
by eqs.~(\ref{eq:PDF-LNAb0}) and (\ref{eq:PDF-LNAb}) in SM \cite{Supp}, the LNA quasi-stationary probability density reads
\begin{eqnarray} 
\label{eq:PDF-LNA-b}
p_{{\rm LNA},\nu,b}^*(N) &=& \phi p_{{\rm LNA},\nu, b}^*(N) + \widetilde{\phi} p_{{\rm LNA},\nu, 0}^*(N).
\end{eqnarray}

Within the LNA, the quasi-stationary average population size is obtained by averaging $N$ over $p_{{\rm LNA},\nu,b}^*(N)$:
\begin{eqnarray} 
\label{eq:av_LNA}
\langle N\rangle^*_{{\rm LNA},\nu,b}= \int_0^{\infty} N p_{{\rm LNA},\nu,b}^*(N) ~dN\,,
\end{eqnarray}
where, it is worth noting, the integral is no longer restricted to a finite support.
As figure \ref{fig:pdmp-predictions}(b) shows, $\langle N\rangle^*_{{\rm LNA},\nu,b}$ is 
as good an approximation of simulation results, as its PDMP counterpart $\langle N\rangle^*_{\nu,b}$ from equation (\ref{eq:Nstarq}).
This is not surprising, and as done in Section \ref{sec:public-good}, it is 
convenient to compute the averages of $N$ using the PDMP approximation,
i.e. by averaging over $p^*_{{\rm PDMP},\nu,b}(N)$  as in eq.~(\ref{eq:Nstarq}).
However, as elaborated below, the LNA via the equation (\ref{eq:PDF-LNA-b}) gives an excellent characterization of the {\it full} $N$-QSD, 
well beyond the scope of the 
PDMP approximation.

\begin{figure}
\begin{center}
\includegraphics{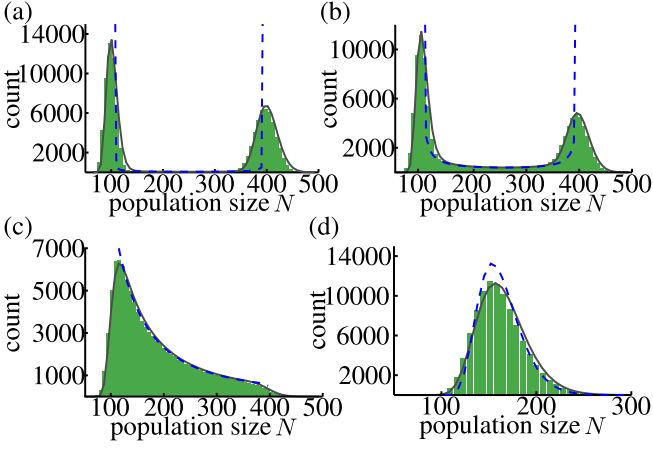}
\end{center}
\caption{Histograms of the population size distribution ($N$-QSD) when $b=0$ (shaded area) compared with 
the predictions of the LNA (solid), from equation~(\ref{eq:PDF-LNAb0}) of the SM~\cite{Supp}, and with the PDMP
predictions (dashed), from $p^*_{\nu,0}$, for different switching rates: 
(a) $\nu=0.01$,  
(b) $\nu=0.1$, (c) $\nu=1$, (d) $\nu=10$, see text. 
Parameters are  $(K_+,K_-,s,x_0)=(400,100,0.02,0.5)$. Here, ${\cal K}=160$.
}
\label{fig:LNAb0}
\end{figure}

\subsection{LNA, $N$-QSD, and noise-induced transitions}
\subsubsection{Pure resource competition scenario, $b=0$}

In the pure resource competition scenario ($b=0$), $p_{{\rm LNA},\nu,0}^*(N)= \pi_0^*(n)/\Omega$  provides an excellent approximation of the $N$-QSD in all switching regimes, as shown in figure \ref{fig:LNAb0}.
In particular, $p_{{\rm LNA},\nu,0}^*$ captures the noise-induced transition arising about $\nu =1$ \cite{HL06,Bena06,KEM1}: When $\nu<1$, 
the switching rate is lower than the population growth rate, and the $N$-QSD and $p_{{\rm LNA},0}^*$ are both bimodal, with peaks at $N\approx K_{\pm}$, see figure \ref{fig:LNAb0} (a,b). 
When  $\nu>1$, the switching rate exceeds the population growth rate, and the $N$-QSD  and $p_{{\rm LNA},\nu,0}^*$ are thus unimodal, with a peak at $N\approx {\cal K}$, see Figure \ref{fig:LNAb0}(c,d).

Figure \ref{fig:LNAb0} also shows that $p_{{\rm LNA},\nu,0}^*(N)$ accurately predicts the peaks, their width and intensity,
and the skewness of the $N$-QSD, whereas the PDMP predictions from equation (\ref{eq:pstarqmar}) only captures the position of the peaks.
This demonstrates how demographic fluctuations, aptly accounted for by the LNA, cause the discrepancies between the $N$-QSD and $p^*_{\nu}$.
\begin{figure}
\begin{center}
\includegraphics{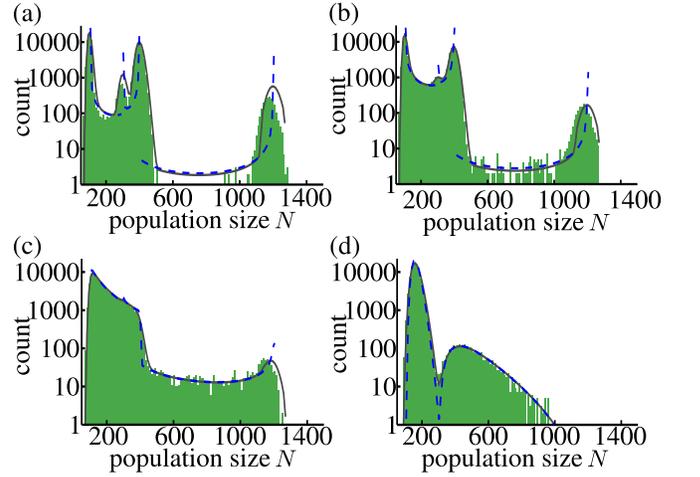} 
\end{center}
\caption{Histograms of the population size distribution ($N$-QSD) when $b=2$ (shaded area) compared with the predictions 
of the LNA (solid), from eq. (\ref{eq:PDF-LNA-b})
and equations (\ref{eq:PDF-LNAb0}) and (\ref{eq:PDF-LNAb}) in the SM \cite{Supp}, and with the PDMP predictions (dashed)
based on eq. (\ref{eq:pstarqmar}), with $q=b$ (when $x=1$) and $q=0$ (when $x=0$), for different switching rates: (a) $\nu=0.01$, (b) $\nu=0.1$, (c) $\nu=1$, (d) $\nu=10$. 
Parameters are  $(K_+,K_-,s,b,x_0)=(400,100,0.02,2,0.5)$.
For the analytical results, we have used the expression (\ref{eq:phi-nu_q}) for $\phi(b)\simeq \phi_{q(b)}$.
}
\label{fig:LNAb}
\end{figure}

\subsubsection{Public-good  scenario, $b>0$}
The LNA expression (\ref{eq:PDF-LNA-b}) also provides an excellent approximation of the $N$-QSD in all switching regimes for the public good scenario ($b>0$), see figure \ref{fig:LNAb}.
In particular, $p_{{\rm LNA},\nu,b}^*$ captures the noise-induced transitions arising about $\nu =1$ and $\nu=1+b$ \cite{KEM1}: 
When $\nu<1$, both conditional population distributions (for fixations to $S$ or $F$) are bimodal, with different peaks.
$N$-QSD and $p_{{\rm LNA},\nu,b}^*$ thus have four peaks at $N\approx  K_{\pm}$ and $N\approx (1+b)K_{\pm}$, see figure \ref{fig:LNAb}(a,b).
When $1<\nu<1+b$, the $S$-conditional distribution is bimodal, whereas the $F$-conditional distribution is unimodal.
The $N$-QSD and $p_{{\rm LNA},\nu,b}^*$ thus have three peaks at $N\approx (1+b)K_{\pm}$ and $N\approx {\cal K}$, see figure \ref{fig:LNAb}(c).
Finally, when $\nu>1+b$, both conditional distributions are unimodal, but with different peaks.
Hence, the $N$-QSD and $p_{{\rm LNA},\nu,b}^*$ are bimodal with peaks at $N\approx {\cal K}$ and $N\approx (1+b){\cal K}$, see figure \ref{fig:LNAb}(d)

As figure \ref{fig:LNAb} shows, $p_{{\rm LNA},\nu,b}^*(N)$  provides a faithful characterization of the $N$-QSD also when $b>0$.
This reiterates that the discrepancies with the PDMP approximation stem from demographic fluctuations.
We also notice that the accuracy of the LNA slightly deteriorates near the lower-intensity peaks at high $N$ and  low $\nu$
(see figure \ref{fig:LNAb}(a)).
These correspond to rare events, usually beyond the scope of the LNA.
Moreover, in those regimes, some assumptions made in the derivation--- {\it e.g.} equation (\ref{eq:approx})---
reach the limit of their validity, see SM \cite{Supp}.

\section{Conclusion}
\label{final}
We have studied the eco-evolutionary dynamics of a population subject to a randomly switching carrying capacity in which one strain has a slight selective advantage over another.
In a model inspired by microbial communities evolving in fluctuating environments, we have considered two scenarios---one of pure resource competition (no interaction between strains) and one in which the slow (cooperating) strain produces a public good---and investigated 
the {\it coupled} effect of demographic and environmental noise.

The population composition has been  characterized  by the fixation probabilities, computed using the 
analytical procedure devised in Ref.~\cite{KEM1}, and, when a public good is produced,   shown to be 
non-trivially correlated with the evolution of the population size. 
As a result, the production of public good gives rise to an \textit{eco-evolutionary game}: 
On the one hand, producing the public good lowers the survival/fixation probability of cooperators; on the other hand, 
it also increases their population size. A social dilemma of sorts therefore ensues and, in a fluctuating environment,
it is a priori not intuitively clear whether there are circumstances under which it is beneficial to produce a public good
and what these conditions may be. 
Since we consider the eco-evolutionary game  in a  population of fixed composition (after fixation)
but whose size fluctuates, we have 
proposed to measure the evolutionary benefit of the public good in terms of the long-term expected number of
individuals of  each strain. 
This is done in the biologically-inspired setting of a metapopulation of non-interacting
communities of varying size composed uniquely by one of the species.
In certain circumstances, that we have determined,
the public good production allows the communities composed of cooperating $S$ individuals to achieve 
a greater long-term increase of their average size than the communities consisting of freeriding $F$ individuals.  
In these conditions, we say that the cooperating strain outcompetes the freeriding one.
We have thus determined, both analytically and with simulations,
the circumstances under which cooperation is beneficial or detrimental to public good producers, 
as well as the conditions under which it is the optimal strategy. Hence, we have demonstrated that 
the rate of switching, along with the selection intensity and the public good parameter, determine when
one species is more successful than another.
Our analysis of the ``eco-evolutionary game'' thus shows that in a fluctuating population 
the evolutionary success of a strain goes beyond having a growth-rate advantage and a higher fixation probability. 

We have also advanced the characterization of the population size distribution by generalizing the linear noise approximation to populations
of fluctuating size, thus accounting for demographic fluctuations about the predictions of the underlying piecewise deterministic Markov process.
While we have found that the linear noise and the piecewise-deterministic Markov process approximations describe the average population size equally well, only the former fully characterizes the population size distribution.
In fact, the linear noise approach accounts for the joint effect of environmental and demographic noise and has allowed us to 
capture the width and skewness of the population size distribution.

This study shows that coupled environmental and demographic noise can greatly influence how the composition and size of a population evolve.
In particular, social interactions between strains---such as public good production---can lead to intricate eco-evolutionary dynamics,
which potentially support cooperation. 
This sheds light on phenomena that are directly relevant to microbial communities, which often feature 
coupled internal and ecological evolution. This can yield the kind of 
 eco-evolutionary game analyzed here,  that can be a potential
 theoretical framework for experimental studies investigating the emergence of cooperative behavior in 
microbial communities of time-fluctuating size.
\\
\begin{acknowledgments}
KW is grateful to the University of Leeds for the hospitality  during the final stage of this work.
EF and KW acknowledge funding by the Deutsche Forschungsgemeinschaft, Priority Programme 1617 
``Phenotypic heterogeneity and sociobiology of bacterial populations'', grant FR 850/11-1,2, and the German Excellence Initiative 
via the program ``Nanosystems Initiative Munich'' (NIM).
MM is grateful for the  support of the Alexander von Humboldt Foundation (Grant No. GBR/1119205 STP) and for 
the hospitality of the University of Munich during the initial phase of this collaboration. 
\end{acknowledgments}

\renewcommand{\thefootnote}{\fnsymbol{footnote}}
\renewcommand{\thefigure}{S\arabic{figure}}
\renewcommand{\theequation}{S\arabic{equation}}
\renewcommand{\thesection}{SM-\Roman{section}}
\setcounter{figure}{0}
\setcounter{equation}{0}
\setcounter{section}{0}

\newpage

\begin{widetext}

\begin{center}
{{\itshape\Large Appendix: Supplementary Material for}\\~\\\Large Eco-Evolutionary Dynamics of a Population with Randomly Switching Carrying Capacity}
\end{center}

In this Supplementary Material, we provide comments concerning the dichotomous noise, notes on the methodology and data avalability,  the
derivation of the probability densities of the piecewise-deterministic Markov process (PDMP), 
complementary results about the mean fixation times, as well as additional discussions about the PDMP  approximation
and the emergence of cooperation in the eco-evolutionary game, and  additional  technical details concerning 
 the linear noise approximation to the population size's quasi-stationary distribution.

\vspace{0.25cm}

In what follows, unless stated otherwise, the notation is the same as in the main text and the equations and
figures refer to those therein. (As in the main text, unless explicitly mentioned otherwise, below we tacitly assume $x_0=1/2$.)

\vspace{0.5cm}

\section{Relationship between dichotomous Markov noise and other forms of environmental noise}
\label{SM1} 

It is worth outlining some of the similarities and differences between the dichotomous Markov Noise (DMN) and the Ornstein-Uhlenbeck process (OUP) that is also commonly used to model environmental noise, see {\it e.g.}~\cite{AMR13}.
Both are {\it colored noises} with exponential auto-correlation functions, see Sec.~II in the main text and Refs.~\cite{HL06,Bena06}.
However, while the Ornstein-Uhlenbeck Process is a Gaussian and unbounded process, the DMN is, in general, \textit{neither}.
In fact, the piecewise-deterministic Markov process (PDMP) \cite{Davis84}
\begin{eqnarray}
	\dot{N}= N\left(1+q-\frac{N}{{\cal K}}\right)+
	\Delta\xi,
	\label{eq:Nflow}
\end{eqnarray}
with the DMN $\xi$, becomes a diffusive process with Gaussian white noise and diffusion constant $D=\Delta^2/\nu$ only in
the limit of $\Delta \to \infty, \nu \to \infty$ and $0<D<\infty$, see, {\it e.g.}, Refs.~\cite{HL06,Bena06,Kithara79}.
The PDMP that we consider in this work has the form:
\begin{eqnarray}
\label{eq:sdebis}
\dot{N}= {\cal F}(N,\xi)=
\begin{cases}
{\cal F}_+(N)  &\mbox{if } \xi=1 \\
{\cal F}_-(N) &\mbox{if } \xi=-1
\end{cases}
\quad \text{with} \quad
{\cal F}_{\pm}(N) \equiv N\left[1+q-\frac{N}{K_{\pm}}\right],
\end{eqnarray}
which coincides with equations (\ref{eq:sde}) of the main text and (\ref{eq:Nflow}) with
$\Delta=(K_+ -K_-)/(2K_+ K_-)$. Since $K_+>K_-\gg 1$, the Gaussian white noise limit is unphysical, and the  PDMP that we consider in this work 
is therefore {\it never diffusive}.

It is also worth noting that, being bounded, the DMN has the great advantage of guaranteeing that the 
fluctuating carrying capacity $K(t)=[(K_+ + K_-)+\xi(t)(K_+ -K_-)]/2$ remains always finite and positive, which would not 
be the case if $\xi(t)$ was given by an OUP.
Furthermore, the DMN can be considered a discrete-step approximation \cite{HL06,Bena06} of
the OUP, but is more mathematically tractable and easier to simulate. 
\section{Notes on methodology \& Data availability}
\label{AppendixB}

Source code for all simulations, resulting data and the \textit{Mathematica} notebook~\cite{Mathematica} used 
for calculations and figures are available electronically~\cite{Supp}.

\subsection{Stochastic simulations}
Using Gillespie's stochastic simulation algorithm (SSA) \cite{Gillespie76}, we have simulated exactly the dynamics described by the master equation (5).
To efficiently ensure that quasi-stationarity was reached \cite{meta}, we have run individual-based simulations until fixation occurred in 
99\% of the realizations (for $\nu\gtrsim s$), or until time reaches $t=10/\nu$ (when $\nu \ll s$).
We have simulated ensembles of $10^4$ realizations of the system, except to determine the various population size distributions 
(for which we used a larger sample of $10^5$ realizations) and when using ``high values'' of $s$ (i.e. for $s={\cal O}(1)$ 
as in figure \ref{fig:delta-phi-s}).
In this case, an even larger sample of $10^6$ realizations was needed to accurately estimate the fixation probability of $S$.

\subsection{Numerical limitations on effective parameter $q(b)$ approach}
To obtain the parameter $q(b)$ used in the formula (15) for $\phi_{q}$, we first recorded the fixation probability from SSA results with 
constant $K=\mathcal{{K}}$, $b\in\{0.1k:\,k\in\mathbb{N},\,k\leq100\}$, and $s\in\{0.02,0.05\}$ ($10^6$ runs each).
For each combination of parameters, we computed $q(b)$ by matching the fixation probability $\phi|_{(1+q)\mathcal{K}}$ of  the 
fitness-dependent Moran model, see equation (\ref{eq:phiNc}), 
with the corresponding fixation probability 
obtained in the  SSA result.

The values of $q(b)$ have then been used to compute $\phi_q$ according to equation~(15) for several values of $\nu$, as shown in figure 3(b).
Due to numerical instabilities in the evaluation of stationary distribution $p^*_{\nu,q}$ in Mathematica~\cite{Mathematica}, numerical evaluations 
of $\phi_q$ occasionally ``failed'' or produced outliers.
Data corresponding to these occasional issues were omitted (without statistical consequences) from our dataset. This has sometimes led to  
some gaps in the lines of the analytical predictions (see {\it e.g.} the green curve in figure 3~(b)).
Furthermore, $q(b)$ has only been determined for a discrete set of $b$ values, which limits the resolution in determining $b_c$ and $b^*$.
Specifically, since the spacing between the values used for $b$ was $0.1$, neither $b^*$ nor $b_c$ has been determined with
an  accuracy higher than $0.1$.
The combination of limited resolution and outliers causes the jaggedness observed  in figure \ref{fig:bstar}(a)
for the graph of $b^*$ obtained by looking for the maximum of equation (21).

\subsection{Data availability: Mathematica notebook \& Linear noise approximation figures 7 and 8}
The direct numerical evaluation of equations (\ref{eq:PDF-LNAb0}) and (\ref{eq:PDF-LNAb}), used to generate the figures 7 and 8,
is commented in the accompanying Mathematica notebook~\cite{Supp}.

\begin{figure}[hbt]
\begin{center}
\includegraphics{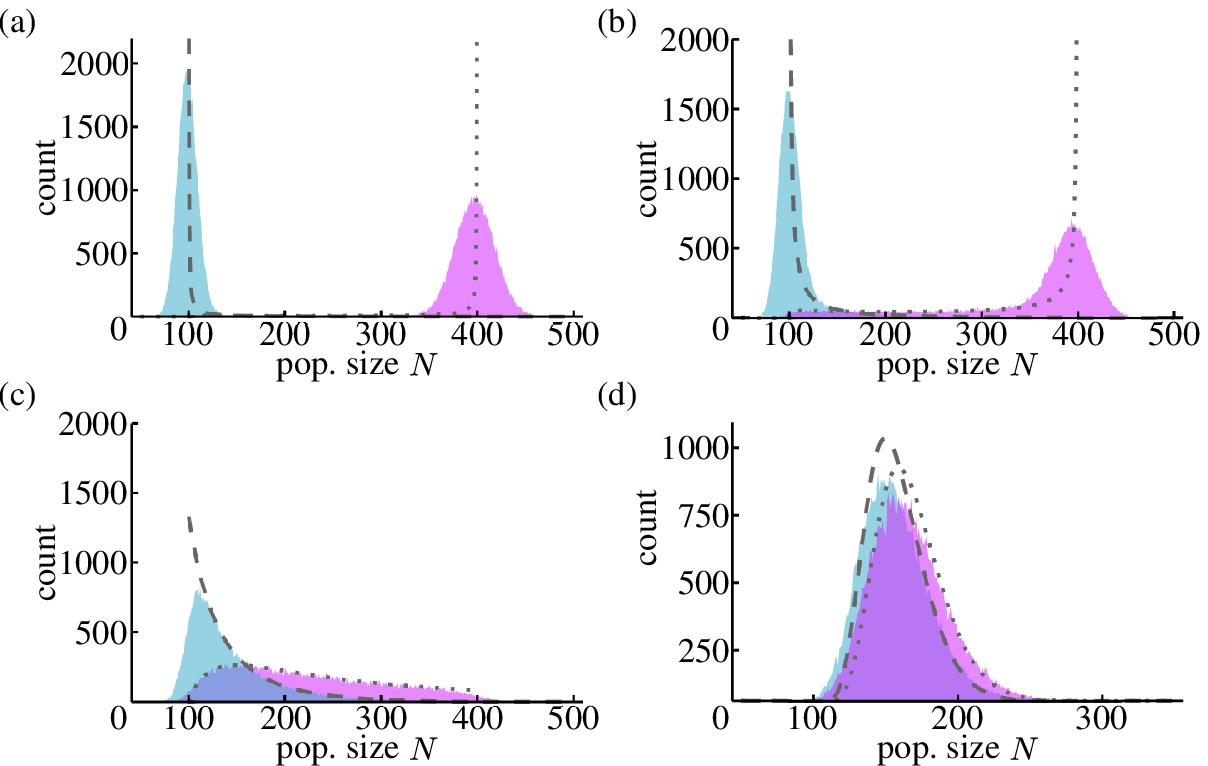}
\end{center}
\caption{
Histograms of population size
($N$-QSD) and  from the joint PDMP PDF (\ref{eq:pstarjoint})  when $b=0$, for (a) $\nu=0.01$, (b) $\nu=0.1$,
(c) $\nu=1$, and (d) $\nu=10$.
Shaded areas correspond to SSA results for $\xi=+1$ (purple) and $\xi=-1$ (cyan);
dashed and dotted lines are from (\ref{eq:pstarjoint}) with $\xi=+1$ and $\xi=-1$, respectively.
Parameters are  $(K_+, K_-,s,x_0,b)=(400, 100,0.02,0.5,0)$
\label{fig:PDMP_joint}}
\end{figure}

\section{Joint and marginal stationary PDFs of the auxiliary PDMP (9)}
\label{AppendixC}
In the main text, we have frequently used of the marginal stationary probability density 
function (PDF) of the single-variate PDMP (9) that reads
\begin{equation}
\label{eq:pstarqmar}
p_{\nu,q}^*(N)=\frac{{\cal Z}_{\nu,q}}{N^2}~\left[\frac{(N_+^*-N)(N-N_-^*)}{N^2}\right]^{\frac{\nu}{1+q}-1},
\end{equation}
and is given by equation (12) in the main text. When $b=q=0$, we have 
$p_{\nu,0}^*(N)\equiv p_{\nu}^*(N)=\frac{{\cal Z}_{\nu}}{N^2}~\left[\frac{(K_+-N)(N-K_-)}{N^2}\right]^{\nu-1}$.

In this section, we outline the derivation of this PDF, as well as that of the joint stationary  probability density 
$p_{\nu,q}^*(N,\xi)$ of  $N$ and $\xi$. For notational simplicity, in the remainder of this section, we write $p_{\nu,q}=p$ and $p^*_{\nu,q}=p^*$. It follows from the Chapman-Kolmogorov equation, 
that $p(N,\xi)$ obeys the master-like equation \cite{HL06}
\begin{equation}
\label{eq:PDF1}
\partial_t p(N,\xi,t) = -\partial_N\left[{\cal F}(N,\xi)p(N,\xi,t) \right] - \nu \left[p(N,\xi,t) - p(N,-\xi,t) \right]\,,
\end{equation}
which can conveniently be rewritten as $\partial_t p(N,\xi,t) = -\partial_N J(N,\xi,t)$ in terms of the probability current~\cite{Hufton16}
\begin{eqnarray}
\label{eq:cur}
J(N,\xi,t)={\cal F}(N,\xi)p(N,\xi,t) 
&+&\nu \int_{N_-^*}^{N}~dN' \left[p(N',\xi,t) - p(N',-\xi,t)\right].
\end{eqnarray}
The first term on the right-hand-side (RHS) of (\ref{eq:cur}) accounts for the probability flowing outside $[N_-^*,N]$ (Liouvillian flow), whereas the second accounts for the random switching.
At stationarity, $\lim_{t \to \infty} p(N,\xi,t)= p^*(N,\xi)$ and $\lim_{t \to \infty} J(N,\xi,t)=J^*(N,\xi)$, with $\partial_t p^*(N,\xi) = -\partial_N J^*(N,\xi)=0$, which implies $\partial_N (J^*(N,\xi) + J^*(N,-\xi))=0$.
With the (natural) zero-current boundary conditions at $N^*_{\pm}$~\cite{HL06}, i.e. $J^*(N,\xi)=0$, we find a simple relationship between the PDFs in each of the environmental states:
\begin{eqnarray}
\label{eq:cur1}
p^*(N,\xi)=-\left(\frac{{\cal F}(N,-\xi)}{{\cal F}(N,\xi)}\right)p^*(N,-\xi).
\end{eqnarray}
With this relation, $\partial_N J^*(N,-\xi)=0$  gives 
\begin{eqnarray}
\label{eq:cur2}
0=\partial_N \left[{\cal F}(N,-\xi)p^*(N,-\xi)\right] + \nu  \left[\frac{1}{{\cal F}(N,-\xi)} + 
\frac{1}{{\cal F}(N,\xi)}\right]\left({\cal F}(N,-\xi)p^*(N,-\xi)\right).\nonumber
\end{eqnarray}
Combined with equation~(\ref{eq:cur1}), this readily yields $p^*(N,\xi)\propto \pm{\mathfrak g}(N)/{\cal F}(N,\xi)$, where
\begin{eqnarray}
\label{eq:g}
{\mathfrak g}(N)={\rm exp}\left[-\nu \int^N dm \left\{
\frac{1}{{\cal F}_{-}(m)}+ \frac{1}{{\cal F}_{+}(m)}
\right\}\right]=\left[\frac{(N_+^*-N)(N-N_-^*)}{N^2}\right]^{\frac{\nu}{1+q}},
\end{eqnarray}
and ${\cal F}_{\pm}$ are defined by eq.~(10).
The {\it joint stationary PDF} giving the probability density of $N$ in each environmental state is thus explicitly given by
\begin{eqnarray}
\label{eq:pstarjoint}
p^*(N,\xi)=\frac{{\cal Z}}{\xi {\cal F}(N,\xi)}
~{\mathfrak g}(N)=\frac{{\cal Z}}{\xi {\cal F}(N,\xi)}~\left[\frac{(N_+^*-N)(N-N_-^*)}{N^2}\right]^{\frac{\nu}{1+q}},
\end{eqnarray}
where ${\cal Z}$ is the normalization constant.
In figure \ref{fig:PDMP_joint}, we compare the predictions of the joint PDF $p^*(N,\xi)$ with the histograms of the population size 
obtained from SSA results, verifying that the PDMP description aptly reproduces the location and number of peaks that characterize the 
quasi-stationary distribution of $N$ (see also \cite{KEM1,SM}).

The marginal stationary PDF $p^*(N)=p^*(N,\xi)+p^*(N,-\xi)$ is thus $p^*(N)\propto [(1/{\cal F}_+(N))-(1/{\cal F}_-(N))]~{\mathfrak g}(N)$, 
which yields the explicit expression  (\ref{eq:pstarqmar}).

It is also useful to notice that, at stationarity, the probability that the PDMP (\ref{eq:sde}) is in the environmental state $\xi$, given a population size $N$ is given by \cite{Hufton16}

\begin{eqnarray}
\label{eq:pcond}
p^*(\xi|N)
=\frac{-\xi {\cal F}(-\xi,N)}{\sum_{\xi=\pm 1} \xi {\cal F}(\xi,N)}.
\end{eqnarray}
\section{Assessment of accuracy of formulas for the fixation probability}
\label{AppendixAssess}
A central point of our analysis is the formula to compute the fixation probability  of $S$ (see section III.C.1 in the main text), 
$\phi$,  with the formula~\cite{KEM1}
\begin{equation}
\label{eq:phi-nubis}
\phi\simeq \int_{K_-}^{K_+} \left(\frac{e^{-N s(1-x_0)}-e^{-N s}}{1-e^{-N s}}\right)~p_{\nu/s}^*(N)~dN,\, 
\quad \text{when $b=0$ \, (no public good production)}.
\end{equation}
When $s = \mathcal{O}(1)$, the assumption of a timescale separation between $N$ and $x$ that underpins the derivation of 
equation (\ref{eq:phi-nubis})  is no longer valid.
As a consequence, the relative deviations between the predictions of eq. (\ref{eq:phi-nu}) and the SSA results for $\phi$ 
increase with $s$, as shown in figure \ref{fig:delta-phi-s}.
To quantify the accuracy of equation (\ref{eq:phi-nubis}), we have compared its predictions with the simulation results 
of $10^6$  realizations obtained for different values of $\nu$ and $s$ spanning between $0$ and $0.25$,
recording the SSA fixation probability $\phi_\textnormal{sim}$.
For each combination of parameters (different colors in Figure \ref{fig:delta-phi-s}), we determined the theoretical prediction
$\phi_\textnormal{th}$ from eq.~(\ref{eq:phi-nubis}) and the percentage deviation between it and the simulation result $\Delta\phi = 100|\phi_\textnormal{th}-\phi_\textnormal{sim}|/\phi_\textnormal{sim}$.
As figure \ref{fig:delta-phi-s} shows, theoretical results reproduce simulations for small $s$, with relative deviations below $10\%$.
Discrepancies increase more and more as the selection intensity is increased towards $s={\cal O}(1)$ (when $s> 0.1$, in 
figure~\ref{fig:delta-phi-s}). The approximation underpinning (\ref{eq:phi-nubis}) is therefore valid in the regime $s\ll 1$, which is 
the regime of weak selection pressure on which we focus  (see main text),
and deteriorates as $s$ approaches $s={\cal O}(1)$.

\begin{figure}[hbt]
\begin{center}
\includegraphics{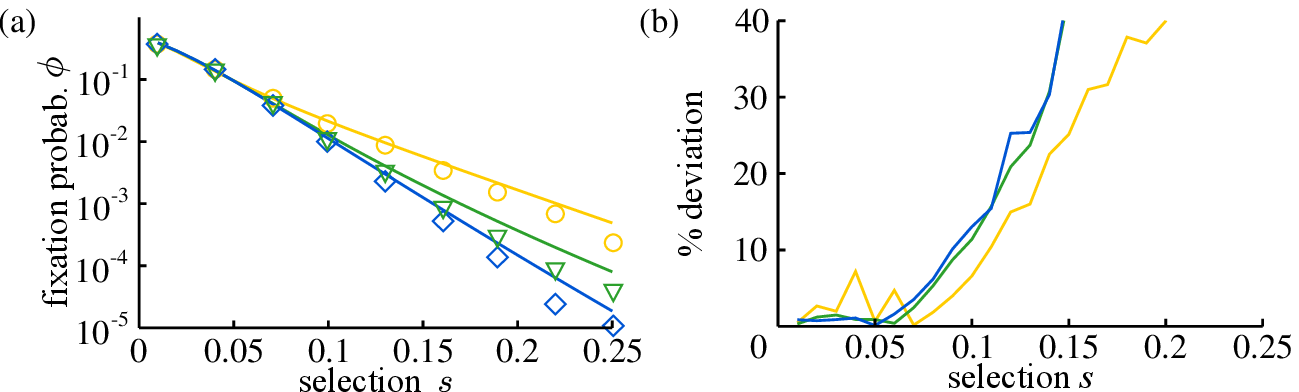}
\end{center}
\caption{
(a) $\phi$ vs. $s$ when $b=0$, with $\nu=0.1$ ($\circ$, yellow),  $\nu=1$ ($\nabla$, green),  $\nu=10$ ($\diamond$, blue).
(b) Accuracy of formula as function of $s$ with $b=0$, measured as the relative deviations from simulation results for $\phi$ 
with the same $\nu$ as in (b), see  text for details.
In both panels, symbols are from simulations ($10^6$ runs in (a)) and solid lines are from eq. (\ref{eq:phi-nu}). Other parameters are $(K_{+},K_-,x_0)=(450,50,0.5)$
} 
\label{fig:delta-phi-s}
\end{figure}

Within the regime $s\ll 1$, we have similarly assessed the accuracy of equation (\ref{eq:phi-nu}) for different switching rates $\nu$.
We simulated $10^4$ realizations of the system, for different values of $s$, and 100 values $\nu$ between 0.001 and 10 and computed the 
percentage deviation $\Delta\phi (\nu)$ as explained above.
Dots in figure \ref{fig:delta-phi-b}(a) are thus based on $\phi_{{\rm sim}}$ obtained by sampling $10^4$ SSA realizations; they
represent the value of $\Delta\phi (\nu)$ recorded at each value of $\nu$, with different colors signaling different values of $s$ (red for $0.05$, cyan for $0.02$).
The dots scatter uniformly, indicating no systematic trend in the deviation.
For $s=0.02$, we observe deviations between $0$ and $8\%$ when $s=0.02$, with an average (solid line) of $2\%$ and standard deviation (shaded area) $\approx2\%$; for $s=0.05$ (red), deviations are between $0$ and $13\%$, with average $4\%$ and standard deviation $2.5\%$.

\begin{figure}[hbt]
\begin{center}
\includegraphics{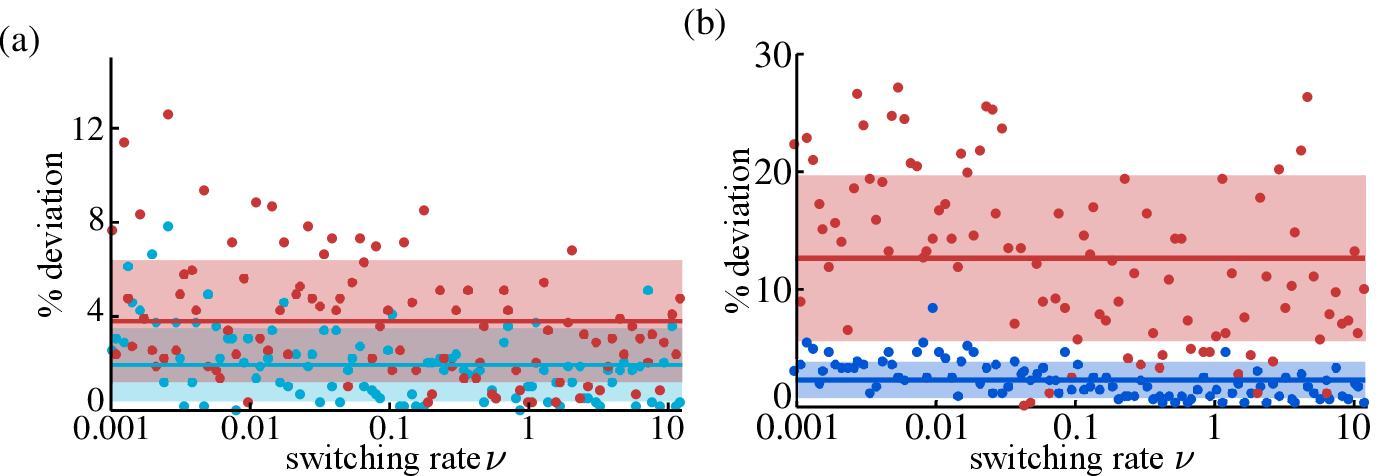}
\end{center}
\caption{
(a) Percentage deviation $\Delta \phi$ between simulation and theory vs. $\nu$, for $b=0$
with $s=0.02$ (cyan) and $s=0.05$ (red).
Dots represent the percentage distance between prediction and simulated value for each $\nu$, solid lines denote the average of the dots, shaded areas the standard deviation around the average.
(b) Same as in panel (a) but 
for $(s,b)=(0.02,0.2)$ (blue) and $(0.05,2)$ (red). 
Other parameters are $(K_{+},K_-,x_0)=(450,50,0.5)$.
\label{fig:delta-phi-b}}
\end{figure}

For the case with public good production, we have used an effective approach and obtained the following expression (\ref{eq:phi-nu_q})
for the fixation probability of $S$ (see section III.C.2) which, with eq.~(\ref{eq:phiNc}), 
 reads~\cite{KEM1}:
\begin{equation}
\label{eq:phi-nu_qbis}
\phi_{q}= \int_{(1+q)K_{-}}^{(1+q)K_{+}}\!\!  \left(\frac{e^{-N s(1-x_0)}-e^{-N s}}{1-e^{-N s}}\right)~
p_{\nu/s,q}^*(N)~dN\,, \quad \text{when $b>0$ \, (public good production)}.
\end{equation}
This expression also builds on a timescale separation between an effective population size and $x$.
Besides the breakdown of the timescale separation when $s = \mathcal{O}(1)$, the accuracy of the approximation
$\phi \simeq \phi_q$ deteriorates for higher values of $b$ and/or $s$, because the fixation of $S$ then becomes 
increasingly unlikely, see figure 3(a), which limits the accuracy with which $q(b)$ is determined and hence the predictions of equation (\ref{eq:phi-nu_q}).

Figure \ref{fig:delta-phi-b}(b) 
shows the results for the percentage deviation of the predictions of equation (\ref{eq:phi-nu_qbis}), 
using the appropriate values of the effective parameter $q(b)$ (excluding a few outliers), 
and $\phi_{{\rm sim}}$ obtained from $10^4$ SSA realizations.
For $s=0.02$, $b=0.2$ (blue), $\Delta\phi$ is between $0$ and $8\%$, with average $3\%$ and a standard deviation of $2\%$.
For $s=0.05$, $b=2$ (red), we observe larger and more scattered $\Delta\phi$ between $0$ and $28\%$, with average $12\%$ and standard deviation $7\%$.
While a deterioration of the  approximation when $s$ and $b$ are increased
can explain the increase in the average $\Delta \phi$,
higher values of $s$ and $b$ also cause lower fixation probabilities for $S$, see figure 3(a).
The corresponding values of $\phi$ are small which results in noisier values of $\phi_{{\rm sim}}$ 
and $\Delta \phi$,  as shown by red data in figure \ref{fig:delta-phi-b}.

\vspace{0.25cm}

Overall, the above analyis  confirms that our approach is able to predict the fixation probability $\phi$ in the regime of weak selection
intensity ($0<s\ll 1$), both when $b=0$ and $0<b={\cal O}(1)$,  with a remarkable accuracy of a few percent  over a  vast range of values $\nu$.

\section{Fixation in the fitness-dependent Moran process \& Mean fixation time under switching carrying capacity}
\label{AppendixD}

\subsection{Fixation in the fitness-dependent Moran process}
To study the fixation properties of the system, we have  used the properties of the fitness-dependent Moran Process (fdMP) outlined in section III.B 
of the main text~\cite{Kimura,Ewens,Antal06,Blythe07}. In a population of large but finite and constant size $N$, the fixation properties  under weak selection of the fdMP  
 can be inferred from the backward Fokker-Planck equation  associated with the generator \cite{Cremer09,Gardiner,vanKampen,Kimura,Ewens,Blythe07}
\begin{eqnarray}
\label{eq:GfdMM}
{\cal G}(x)|_{N}=
g(x)~\frac{x(1-x)}{N}\left[-Ns\frac{d}{dx}+\frac{d^2}{dx^2}\right], \quad \text{where } g(x)=1+bx.
\end{eqnarray}
For an initial fraction $x_0$ of $S$ individuals, the fixation probability $\phi(x_0)|_N$ of $S$  
obeys ${\cal G}(x)|_{N}\phi(x)|_{N}=0$, with $\phi(1)|_{N}=1$ and $\phi(0)|_{N}=0$ (absorbing boundaries at $x=0,1$). Yielding the result
\begin{eqnarray}
\phi(x_0)|_{N}&=&\frac{e^{-N s(1-x_0)}-e^{-N s}}{1-e^{-N s}}\,,
\label{eq:phiNc}
\end{eqnarray} 
given as equation (\ref{eq:phiNc}) in the main text.

The generator (\ref{eq:GfdMM}) can also be used to study when fixation occurs in 
the fdMP in the realm of the diffusion approximation ~\cite{Kimura,Ewens,Blythe07,Cremer09}. Quantities of  particular interest, 
are the unconditional mean fixation time (MFT)--- which is the average time to reach any of the absorbing states, here either $x=0$ or $x=1$---as well as the conditional MFTs---the mean 
time to reach a specific absorbing boundary. The unconditional MFT is obtained by solving ${\cal G}(x_0)|_{N}~T(x_0)|_{N}=-1$ subject to 
$T(0)|_N=T(1)|_N=0$~\cite{Gardiner,Kimura,Cremer09}. The conditional MFT to reach $x=1$ is denoted by $T^S(x_0)|_{N}$, while $T^F(x_0)|_{N}$ 
is the (conditional) MFT conditioned to  reach $x=0$. The MFTs and the fixation probabilities are related by $T(x)|_{N}=\phi T^S(x_0)|_{N} + (1-\phi) T^F(x_0)|_{N}$.
Explicit, but unwieldy, expressions for the MFTs in the fdMP can be obtained \cite{Gardiner,Kimura,Cremer09,Ewens}, {\it e.g.} 
 the unconditional MFT in the case $b=0$ reads
\begin{eqnarray}
\label{eq:MFTb0expl}
T(1/2)|_N &=& \frac{1}{s} \left\{ (1-2 \phi(1/2)|_N) (\ln(Ns) +\gamma) +
\mathrm{e}^{-\frac{N s}{2}} \mathrm{Ei}\left(\frac{Ns}{2}\right) -
\mathrm{e}^{\frac{N s}{2}} \mathrm{Ei}\left(-\frac{Ns}{2}\right)
 \right.\nonumber\\
&+&\left.\phantom{\frac{N}{s}}
\mathrm{e}^{N s} \phi(1/2)|_N \mathrm{Ei}(-Ns)-\mathrm{e}^{-N s} (1 - \phi(1/2)|_N) \mathrm{Ei}(Ns)
\right\}\label{eq:fdmPT}\,,
\end{eqnarray}
where $\gamma\approx0.577...$ is the Euler-Mascheroni constant, and $\mathrm{Ei}(z)=\int_{-z}^{\infty}dz \frac{e^{-z}}{z}$ is 
the exponential integral. Hence, in the 
regime where $s\ll 1$, with $Ns \gg 1$, and $s(\ln{N})\ll 1$ (with $x_0$ is sufficiently separated from $x=0,1$), 
$T(x_0)|_{N} \sim (\ln{N})/s$~\cite{Kimura,Ewens,Blythe07},  with a subleading prefactor $\sim \ln{N}$.
The conditional MFTs exhibit the same behavior $T^{S/F}(x_0)\sim T(x_0)={\cal O}(1/s)$ to leading order when 
$s\ll 1$, see figure~\ref{fig:fdmp-MFT}. 
A similar behavior also holds 
when $b>0$ and $s\ll 1$, with  a subleading prefactor that then depends (weakly) 
on the public good parameter $b={\cal O}(1)$, specifically
 $T(x_0)|_{N} \sim (\ln{(N-{\cal O}(b))})/s$, as confirmed by figure \ref{fig:fdmp-MFT}. The public good parameter $b$, in fact, reduces the relaxation time of 
 $x$, see equation (7), which results in a weak reduction of the unconditional and MFTs with respect to the case with 
 $b=0$, see also figure \ref{fig:fdmp-MFT}(b,c). The most relevant
 point for our purposes, is the fact that the unconditional MFT  of 
the fdMP scales as ${\cal O}(1/s)$ to leading order when $s\ll 1$ and $Ns\gg 1$, and so do the conditional MFTs, 
in both  cases $b=0$ and $b>0$.

\begin{figure}
\begin{center}
\includegraphics[width=\linewidth]{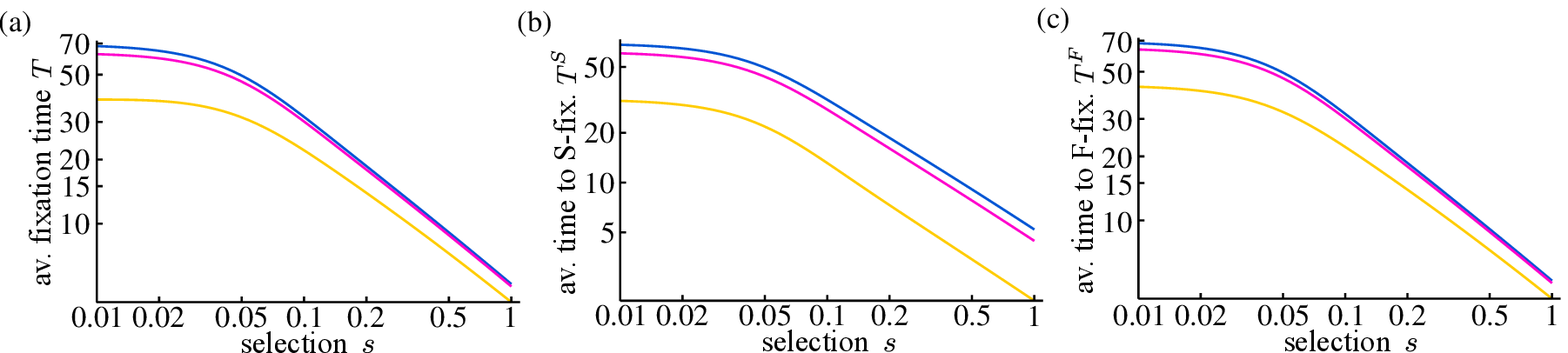}
\end{center}
\caption{
(a) MFT $T|_{N}$ vs. $s$ for the fdMP, given by $T|_N=\phi|_N T^S|_N+(1-\phi|_N)T^F|_N$. 
In the case $b=0$ (blue), this corresponds to equation (\ref{eq:MFTb0}).
(b) $T^S|_{N}$ vs. $s$ for the fdMP, from the solution of the appropriate equation associated with the generator (\ref{eq:GfdMM}).
(c) $T^F|_{N}$ vs. $s$ for the fdMP, obtained as $T^S$.
The population in the fdMP is of  constant size $N=100$ and the  effect of the public good parameter $b$ 
is to reduce the relaxation time of $x$, and thus to lower all the MFTs with respect to the case $b=0$. However, the MFTs
always scale as ${\cal O}(1/s)$ to leading order when $s\ll 1$.
In all panels,  and $b=0$ (blue), $b=0.2$ (pink), $b=2$ (yellow), $x_0=0.5$.
\label{fig:fdmp-MFT}}
\end{figure}

Results (\ref{eq:GfdMM})-(\ref{eq:MFTb0expl}) can be used to obtain the fixation properties of
fdMP in a population of finite size subject to a {\it constant} carrying capacity $K$. This is done  
by setting $N=K$ in (\ref{eq:GfdMM})-(\ref{eq:MFTb0expl}), which 
neglects the rare/unlikely early fixation events  occurring during the 
exponential growth phase (assuming $x_0=1/2$).

\subsection{Mean fixation times with a switching carrying capacity}

In a population subject to a randomly switching carrying capacity, with no public good production, the size and growth rate are independent of the composition.
As explained in Ref.~\cite{KEM1,SM}, when $b=0$, the conditional and unconditional MFTs admit the same scaling to leading order when $s\ll 1$, i.e. 
$T^{S/F}(x_0)\sim T(x_0)={\cal O}(1/s)$, and we 
 can  use the  approach outlined in Section III.C.1, to compute~\cite{SM}
\begin{eqnarray}
 \label{eq:MFTb0}
T(x_0)\simeq \int_{K_-}^{K_+} T(x_0)|_N p_{\nu/s}^*(N)~dN.
\end{eqnarray}
Figure \ref{fig:MFT}(a) shows that the predictions of this  formula (blue line)
agree extremely well with SSA results ($\diamond$). 
This confirms that under weak selection and $b=0$, the unconditional MFT scales as in the fdMP, i.e. 
$T(x_0) ={\cal O}(1/s)$ when $s \ll 1$ and $\langle K \rangle s\gg 1$. 
This implies that after $t\gtrsim 1/s$ fixation is likely to have occurred, and that the population size is at quasi-stationarity when $t\gg 1/s$.
Quite remarkably, we also notice in figure \ref{fig:MFT}(a) that even when  $s={\cal O}(1)$ there is a good 
 agreement between the predictions of eq.~(\ref{eq:MFTb0}) and SSA results.

\begin{figure}[hbt]
\begin{center}
\includegraphics[width=\linewidth]{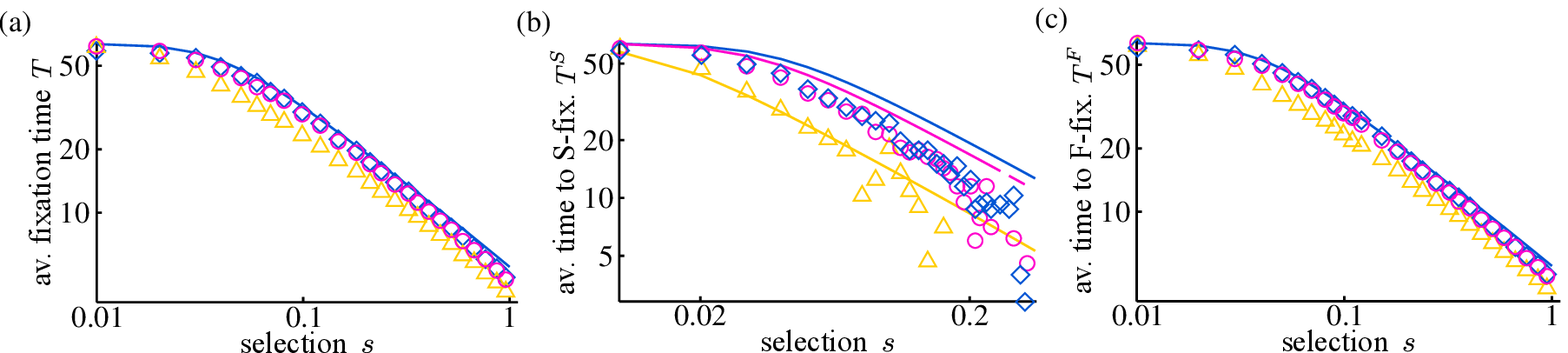}
\end{center}
\caption{
(a) MFT $T$ vs $s$.
Symbols are from simulations ($10^4$ realizations) and the solid line shows $T(x_0)$
given by eq.~(\ref{eq:MFTb0}) in the case $b=0$.
(b) $T^S$ vs $s$.
Solid lines are the results of equation (\ref{eq:TSapprox}) .
(c) $T^F$ vs $s$. 
The solid line is the result of equation (\ref{eq:TFapprox}).
In all panels, $\nu=0.1$ and $b=0$ (blue, $\diamond$), $b=0.2$ (pink, $\circ$), $b=2$ (yellow, $\triangle$), 
other parameters are $(K_-, K_+, x_0)=(50,450,0.5)$.
\label{fig:MFT}
}
\end{figure}

\begin{figure}[hbt]
\begin{center}
\includegraphics[width=\linewidth]{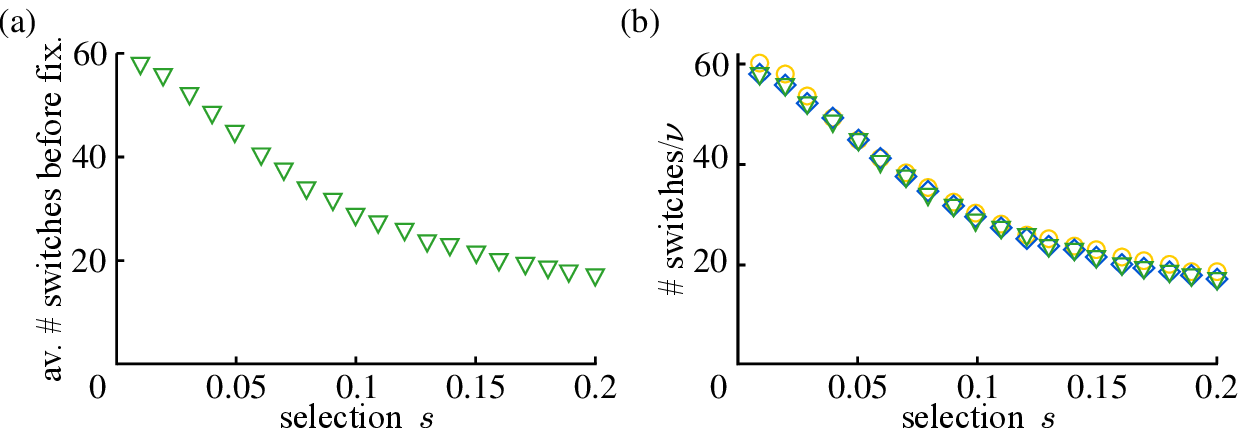}
\end{center}
\caption{
(a) Average number of switches prior to fixation vs. $s$ for $\nu=1$ is  shown
to be of order ${\cal O}(\nu/s)$ when $s\ll 1$.
(b)  Average number of switches prior to fixation rescaled by a factor $1/\nu$ vs. $s$ for $\nu=0.1$
(circles), $\nu=1$ (downside triangles) and $\nu=10$ (diamonds).
In both panels, simulation results are averaged over $10^4$ realizations.
Parameters are $(K_+,K_-,b)=(450,50,0)$. In panel (b), the different symbols essentially collapse onto a single curve
showing that, on average, the number of switches prior to fixation increase linearly with $\nu$
and mostly independent of the population size, see text.
\label{fig:avs}
}
\end{figure}

In the case $b>0$, the evolution of the population size and its composition  are coupled.
As discussed in the main text, $S$ is less likely to fixate when $b$ is increased, 
and the population size at fixation  depends on which species takes over (the population size is typically larger when $S$ fixates). 
On the other hand, according to eq.~(7), increasing $b$ reduces the relaxation time of  $x$. Since these two effects 
balance each other, we expect the effect of $b>0$ to be even weaker in the switching environment than in the fdMP
with constant population size. Having seen  that $b$ has a weak effect on the MFTs, we  anticipate 
that the MFTs with a switching carrying capacity exhibit a similar behavior as those of the fdMP. To verify this
picture and figure out 
on which timescale fixation occurs when $b>0$, we have considered the fixation of $S$ and $F$ separately by studying their 
conditional MFTs~\cite{SM}. For this, we can attempt to generalize the approach used in the case $b=0$
and consider the averages over the conditional stationary PDFs $p^*_{\nu/s,b}(N)$ and $p^*_{\nu/s}(N)$ obtained 
from (\ref{eq:pstarqmar}) with $q=b$ and $q=0$, i.e. 
\begin{eqnarray}
T^S (x_0)&\simeq& \int_{(1+b)K_-}^{(1+b)K_+}T^S(x_0)|_N p^*_{\nu/s,b}(N)dN\,,\label{eq:TSapprox}\\
\label{eq:TFapprox}
T^F (x_0)&\simeq & \int_{K_-}^{K_+}T^F(x_0)|_N p^*_{\nu/s}(N)dN\,,\\
T(x_0)&\simeq&\phi_{q(b)} T^S(x_0) + (1-\phi_{q(b)}) T^F(x_0)\,, \label{eq:MFT}
\end{eqnarray}
where $T^{S/F}(x_0)|_N$ are the conditional MFTs of the fdMP with $b>0$. 
A clear limitation of formula (\ref{eq:TSapprox}) and (\ref{eq:TFapprox}) stems from the fact that 
$p^*_{\nu/s,b}(N)$ and $p^*_{\nu/s}(N)$
provide a good approximation of the $N$-QSD in the  quasi-stationary state that is  reached after $t\gg 1/s$,
see Ref.~\cite{videos}, i.e. well after fixation has typically occurred.
However, since the population size and the parameter $b$ 
only yield subleading contributions to the MFTs of the fdMP when $s\ll 1$, we expect that formula (\ref{eq:TSapprox})-(\ref{eq:MFT})
are still able to capture how the MFTs scale to leading order under weak selection intensity.
The comparison of SSA results for the MFTs with $b>0$ reported in  figure \ref{fig:MFT},
and their comparison with those of figure (\ref{fig:fdmp-MFT}) confirm this picture. 
Since the $b$-dependence of the MFTs  in figure (\ref{fig:fdmp-MFT})  is clearly subleading,
we can simplify the evaluation of (\ref{eq:TSapprox})
by setting $g\equiv 1+b$ (and similarly $g\equiv 1$ in (\ref{eq:TFapprox})). As shown in  figure (\ref{fig:MFT})(b),
this does not affect the leading behavior of $T^S$.

Figures \ref{fig:MFT}(b,c) show that the simplified formula (\ref{eq:TSapprox}) and (\ref{eq:TFapprox}) indeed 
correctly predict that the conditional MFTs scale as ${\cal O}(1/s)$
 when $s\ll 1$, even if, as expected, they overstimate the SSA results for $T^S$ and $T^F$. Hence, equation (\ref{eq:MFT}) predicts that to leading order 
the uncoditional MFT scales as ${\cal O}(1/s)$, which is in good agreement with the SSA results reported in figure 
\ref{fig:MFT}(a).
As $s$ and $b$ increase, the fixation of $S$ becomes less  likely and thus   $T(x_0)\simeq T^F(x_0)$, as shown by  figures \ref{fig:MFT}(a)
and (c). We notice that SSA results reported in figure \ref{fig:MFT} confirm that the MFTs with randomly switching 
carrying capacity depend even more weakly on the public good 
parameter $b$ than in the fdMP where $N$ is constant.

We therefore conclude that, under weak selection $1/\langle K \rangle \ll s\ll 1$, and with $b={\cal O}(1)$, the MFTs in the case $b>0$
 scale as ${\cal O}(1/s)$. This means that  the fixation in the public good scenario ($b>0$) is likely to have 
 occurred when $t\gtrsim 1/s$, as in the case $b=0$.  This also implies that, when $b\geq 0$, the
population size is most probably at quasi-stationarity when $t\gg 1/s$,
and its composition consists then of only $F$ or $S$ individuals.
 These results thus indicate that, to leading order in $1/s$ (with $\langle K \rangle s\ll 1$),
the MFTs here scale as in the absence of external noise. 
Hence, while environmental noise has a significant
effect on the fixation probability (see Section III.C in the main text), its effect on the MFTs is  much less important,
as captured by the formula (\ref{eq:TSapprox})-(\ref{eq:MFT}).
A consequence of these results is that 
 the population experiences, on average, ${\cal O}(\nu/s)$
switches prior to fixation  when $b\geq 0$ and  $1/\langle K \rangle \ll s\ll 1$. In fact, the figure \ref{fig:avs}
confirms that the average number of switches prior to fixation scales as $1/s$ to leading order when $1/\langle K \rangle \ll s\ll 1$ 
in the case $b=0$. 
In figure \ref{fig:avs}(b), we show that the average number of switches prior to fixation increases 
linearly with $\nu$, with simulation data for different values of $\nu$  essentially collapsing when 
rescaled by a factor $1/\nu$.  
Since the population size greatly varies when $\nu$ changes, see, {\it e.g.}  the videos in  
\cite{videos}, the fact that the average number of switches increases simply linearly with $\nu$
is a strong indication of its weak dependence on the population size, and is a further argument supporting the rescaling 
$\nu \to \nu/s$ in formula (\ref{eq:phi-nu}) and (\ref{eq:phi-nu_q}).

\section{Supplementary information on the PDMP approximation and the ``Eco-evolutionary game''}
\label{AppendixPDMP}
\subsection{PDMP approximation and average number of individuals}
The analysis of  the correlations between population size and its composition (Section IV.A), and that of the 
``eco-evolutionary game'' (Section IV.B), relies largely on properties of the average 
population size at quasi-stationarity given by 
\begin{equation}
\label{eq:NstarDef}
\langle N \rangle_{\nu,b}^*=\phi_{b}\langle N \rangle_{\nu,b}^*+(1-\phi_{b})\langle N \rangle_{\nu,0}^*
=(1+b)
\phi_b\langle N   \rangle_{\frac{\nu}{1+b},0}^* + (1-\phi_{b})\langle N   \rangle_{\nu,0}^*\,,
\end{equation}
where $\phi_b$
is the fixation probability of species $S$ under a public good parameter $b$, 
within what in the main text is referred to as the ``PDMP approximation''. This approximation consists of 
averaging the population size $N$ over the marginal PDF (\ref{eq:pstarqmar}) of the PDMP (\ref{eq:sde}).

To derive equation (\ref{eq:NstarDef}), which coincides with equation (\ref{eq:Nstarq}) of the main text, we first notice that $\langle N \rangle_{\nu,b}^*$ consists of
the average population size conditioned to the fixation of $F$ and $S$, i.e.
$\langle N \rangle^*_{\nu,b}=\langle N_F \rangle^*_{\nu,b}+\langle N_S \rangle^*_{\nu,b}$.
The fixation of $F$ occurs with probability $\widetilde{\phi}_b=1-\phi_b$, and results in a global growth rate $g=1$, yielding
\begin{eqnarray}
\label{eq:NF}
 \langle N_F \rangle^*_{\nu,b}=\langle N | x=0 \rangle^*_{\nu,b}=\widetilde{\phi}_b \langle N \rangle^*_{\nu,0}=
\widetilde{\phi}_b \int_{K_-}^{K_+} N p_{\nu}^* (N) dN,
\end{eqnarray}
where $\langle N \rangle^*_{\nu,0}$ is the quasi-stationary average population size when $b=0$
and the integration is over $p_{\nu}^*\equiv p_{\nu,0}^*$ given by eq.~(\ref{eq:pstarqmar}). 
Similarly, the fixation of $S$ occurs with probability $\phi_b$, after which $g=1+b$, yielding
\begin{eqnarray}
\label{eq:NS}
 \langle N_S \rangle^*_{\nu,b}=\langle N | x=1 \rangle^*_{\nu,b}=\phi_b \langle N \rangle^*_{\nu,b}
=\phi_b \int_{(1+b)K_-}^{(1+b)K_+} 
N p_{\nu,b}^* (N) dN
=(1+b)\phi_b \langle N \rangle^*_{\frac{\nu}{1+b},0}.
\end{eqnarray}
The last equality is obtained by performing the change of variable $N \to N/(1+b)$ and allows us to express $\langle N \rangle^*_{\nu,b}$ in 
terms of the average when $b=0$.
Putting everything together, we obtain eq.~(\ref{eq:Nstarq}):
\begin{eqnarray}
\label{eq:Nstarqbis}
\langle N \rangle_{\nu,b}^*&=& (1+b) \phi_b\langle N \rangle_{\frac{\nu}{1+b},0}^* + (1-\phi_b)\langle N \rangle_{\nu,0}^*\nonumber\\
&\simeq&(1+b) \phi_{q(b)}\int_{K_-}^{K_+}\!\!\!\! N p_{\frac{\nu}{1+b}}^*(N)~dN
+ (1-\phi_{q(b)})\int_{K_-}^{K_+} N p_{\nu}^*(N)~dN, 
\end{eqnarray}
where in the last line we have used the approximation $\phi_b \simeq \phi_{q(b)}$ given by equation (\ref{eq:phi-nu_q}).
Figure 3(b) shows that predictions of $\langle N\rangle_{\nu,b}^*$ obtained with this approach are as close to simulation 
results as their counterparts obtained by averaging over the PDF obtained within the
linear noise approximation of section V (see also section \ref{AppendixE} below). It is clear from (\ref{eq:phi-nu_q})
that $\langle N \rangle^*_{\nu,b}$ is an increasing function of $b$ since $\langle N \rangle^*_{\nu,0}$ is a decreasing function 
of $\nu$~\cite{comment}.

In Section IV, we have often considered the limiting regimes of very fast/slow switching, $\nu \to \infty,0$,
in which the analytical formula greatly simplify. To obtain these simplified expressions,
it suffices to notice that 
\begin{eqnarray*}
\int_{K_-}^{K_+} N p_{\nu}^* (N) dN=
\begin{cases}
{\cal K}  &\mbox{when } \nu \to \infty \\
\langle K \rangle &\mbox{when } \nu \to 0
\end{cases}
\end{eqnarray*}
Hence,  when $\nu\gg 1$, we have $\langle N_F \rangle^*_{\nu,b}\to (1-\phi_b){\cal K}$ and
 $\langle N_S \rangle^*_{\nu,b}\to (1+b)\phi_b{\cal K}$. Similarly, when $\nu\ll s$, we have
 $\langle N_F \rangle^*_{\nu,b}\to(1-\phi_b)\langle  K \rangle $ and
 $\langle N_S \rangle^*_{\nu,b}\to(1+b)\phi_b\langle  K \rangle$. Hence, from (\ref{eq:Nstarq}) and using $\phi\simeq \phi_{q(b)}$
 we obtain the average population size in the limiting regimes:
\begin{eqnarray}
\label{eq:Nas}
\langle N \rangle^*_{\nu,b}=
\begin{cases}
\left[1+b \phi_{q(b)}^{(\infty)}\right]{\cal K} &\mbox{when } \nu \to \infty \\
\left[1+b\phi_{q(b)}^{(0)}\right]{\langle K \rangle} &\mbox{when } \nu \to 0.
\end{cases}
\end{eqnarray}
The limiting behavior reported as dashed lines in figures 3(b) and 4(b) can readily be obtained from equations~(\ref{eq:Nas}).

\subsection{Best conditions for cooperation in the eco-evolutionary game}

A finite well-mixed population of constant size is the natural setting of evolutionary game theory (EGT).
The notion of evolutionary stability is central to EGT since an evolutionary stable strategy, when adopted by a population, cannot be invaded and replaced by an alternative strategy.
For a population with two possible strategies, one is evolutionary stable if it satisfies the so-called invasion and replacement conditions 
\cite{Nowak,Broom13}.
As a result, the sole fact that one strategy has a higher fitness than another does not guarantee that it is
evolutionary  stable since an individual of the other type may have a better chance to fixate the population.

For the model considered here, in a finite and static population, the strain $F$ has always a higher fitness than $S$, and the fixation probability of $S$ vanishes exponentially with the population size, see equation (\ref{eq:phiNc}).
In a finite and static population, $F$ is therefore evolutionary stable, and in this sense always superior to $S$.

\vspace{0.25cm}

The situation is radically different in the eco-evolutionary game considered here since the population continues to evolve
in a {\it fluctuating environment} even after fixation, and the notions of non-invadibility / non-replacement are no longer suitable
to measure the species evolutionary success:

As discussed in Section IV.B of the main text, even if $S$ has always a lower fitness and a lesser chance to fixate than $F$, its occasional
fixation can prove very rewarding since it allows cooperators to establish a large community of $S$ individuals (of a  size that
can be significantly larger than the size of an average community of $F$ individuals).
In the context of the interpretation of the  eco-evolutionary game in terms of
a biologically motivated metapopulation of non-interacting communities, 
see Section IV.B of the main text, we have  proposed to measure the success of $S$ and $F$ in this eco-evolutionary game 
by computing the difference 
between the expected long-term number of individuals $\Delta S_{\nu,b}$ and $\Delta F_{\nu,b}$, compared to the $b=0$ case.
$\Delta S_{\nu,b}$ and $\Delta F_{\nu,b}$ thus serve as expected payoffs in our eco-evolutionary game.
In the PDMP approximation, we can use our effective approach (see section III.C.2) and equations (\ref{eq:NF}) and (\ref{eq:NS}), to obtain
\begin{eqnarray}
\Delta S_{\nu, b}&=& (1+b)\phi_{q(b)}\int_{K_-}^{K_+}  N p_{\frac{\nu}{1+b}}^*(N)~dN-\phi_0\int_{K_-}^{K_+}  N p_{\nu}^*(N)~dN\label{eq:DS}\\
\Delta F_{\nu,b}&=& (\phi_0 - \phi_{q(b)})\int_{K_-}^{K_+}  N p_{\nu}^*(N)~dN\,,\label{eq:DF}
\end{eqnarray}
where $p^*_\nu(N)$ is readily obtained by setting $q=0$ in equation (\ref{eq:pstarqmar}). 
In the limiting regimes $\nu \to \infty,0$, with (\ref{eq:Nas}), we obtain
\begin{eqnarray}
\label{eq:}
\Delta S_{\nu, b}=
\begin{cases}
\left[(1+b) \phi_{q(b)}^{(\infty)}-  \phi_{0}^{(\infty)}\right]{\cal K} &\mbox{when } \nu \to \infty \\
\left[(1+b) \phi_{q(b)}^{(0)}-  \phi_{0}^{(0)}\right]{\langle K \rangle} &\mbox{when } \nu \to 0.
\end{cases}
\quad
\text{and} \quad
\Delta F_{\nu, b}=
\begin{cases}
\left[ \phi_{0}^{(\infty)} - \phi_{q(b)}^{(\infty)}\right]{\cal K} &\mbox{when } \nu \to \infty \\
\left[ \phi_{0}^{(0)} - \phi_{q(b)}^{(0)}\right]{\langle K \rangle} &\mbox{when } \nu \to 0.
\end{cases}
\end{eqnarray}

As shown in figure 5,  $\Delta S_{\nu, b}$ is non-monotonic in $b$ and has a maximum for $b=b^*$.
This is then the optimal value of $b$ for the cooperating strain $S$ (given $s$, $\nu$, $K_\pm$).
Figure \ref{fig:bstar}(a) shows the dependence of the optimal value $b^*=b^*(\nu,s)$ on $\nu$, for different intensities of the selection pressure $s$.
Clearly, $b^*=b^*(\nu,s)$  exhibits a complex, non-monotonic, dependence on $\nu$ and  decreases when $s$ increases, in 
a similar fashion to $b_c$ (see main text).
In figure~\ref{fig:bstar}(a), symbols are from simulations and the lines have been obtained from evaluating the 
maximum of equation (\ref{eq:DS}).

\begin{figure}
\begin{center}
\includegraphics{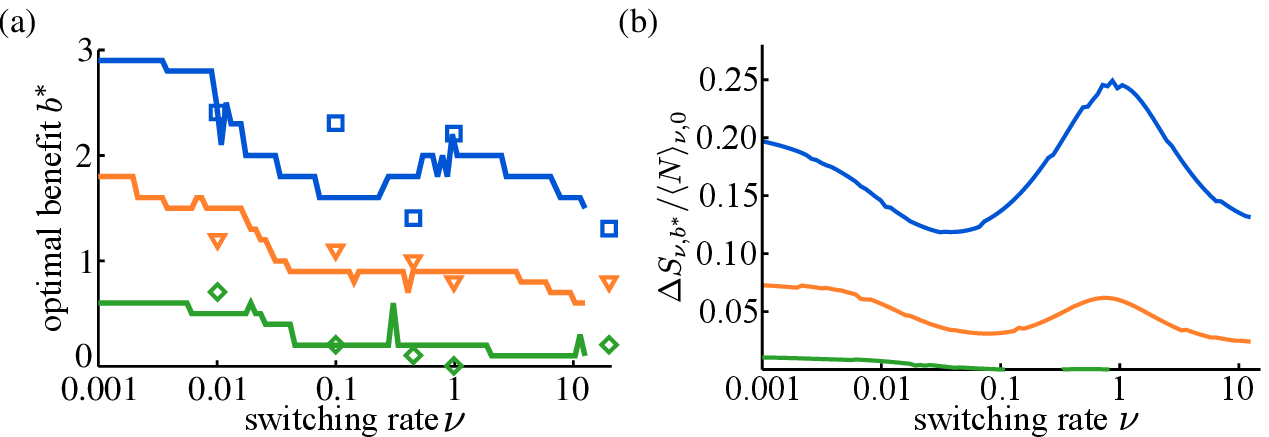}
\end{center}
\caption{(a) 
Optimal public good benefit parameter for the cooperating $b^*$ vs $\nu$ for $s=0.02$ (blue), $s=0.03$ (orange), and $s=0.05$ (green).
Symbols are results from simulations and solid lines are from (25)
(b) $\Delta S_{\nu,b^*}/\langle N   \rangle_{\nu,0}^*$ vs. $\nu$, obtained gives the highest payoff received by $S$ by producing the 
public good at optimal value $b=b^*(\nu,s)$ obtained from (25)
for $s=0.02$ (blue), $s=0.03$ (orange), $s=0.05$ (green), see below
and main text.
Other parameters are $(K_+,K_-,x_0)=(450,50,0.5)$.
}
\label{fig:bstar}
\end{figure}

Figure \ref{fig:bstar}(b) shows $\Delta S_{\nu, b^*}/\langle N\rangle_{\nu,0}^*$: the optimal payoff for cooperators divided by the long-time 
average population at $b=0$.
In other words, it shows how much bigger is, on average, the best-performing cooperating population, compared to 
the average population at $b=0$.
For sufficiently low $s$, e.g. for $s=0.02$ (blue), the public good can make the average number of $S$ individuals be up to $12\%-25\%$ larger than the average population at $b=0$ ($\Delta S_{\nu,b^*}/\langle N \rangle_{\nu,0}^*\approx 0.12-0.25$ across all values of $\nu$).
Figure \ref{fig:bstar}(b) corresponds to results at quasi-stationarity, i.e. after fixation has occurred (with a smaller probability for $S$ than
$F$) and  therefore shows the actual long-term eco-evolutionary payoff for cooperation: In the optimal conditions, the $S$ strain can gain a significant benefit from the production of a public good.

As discussed in Section IV.B of the main text, there are conditions under which $S$ receives a higher expected payoff than $F$ in
the sense that $\Delta S_{\nu,b}> \Delta F_{\nu,b}$. When this happens, cooperating 
is not only beneficial but  is also advantageous for $S$.
We have considered that for given parameters $(\nu,s)$, it is best to cooperate for the production of a public good with benefit parameter $b$ when the following two conditions are satisfied:
(a) $\Delta S_{\nu,b}> \Delta F_{\nu,b}$;
(b) $b=b^*(\nu,s)<\beta(\nu,s)$.
These conditions ensure (a) that $S$ receives a higher payoff than $F$, and (b) that $S$ receives the maximum payoff under the switching rate $\nu$ and selection strength $s$.
On the other hand, species $F$ always outperforms $S$ when $b>\beta(\nu,s)$ since it then receives a higher expected  payoff 
than $S$, with $\Delta F_{\nu,b}$
that is an increasing function of $b$ for all values of $\nu$ and $s$, see figure 5(c).

As shown in figure 6(b) the phases (ii), ($0<\Delta S_{\nu,b}< \Delta F_{\nu,b}$) and (iii) ($\Delta S_{\nu,b}> \Delta F_{\nu,b}$) 
are separated by the value $b=\beta(\nu,s)$ at which $\Delta S_{\nu,\beta}=\Delta F_{\nu,\beta}$, defined as by the solution of
\begin{eqnarray}
 \label{eq:beta}
\hspace{-3mm}
\frac{1}{1+\beta}\left(\frac{2\phi_0}{\phi_{q(\beta)}}-1\right)= \frac{\int_{K_-}^{K_+}N p^*_{\frac{\nu}{1+\beta}}~dN}{\int_{K_-}^{K_+}N p^*_{\nu}~dN}\,.
\end{eqnarray}

It is noteworthy that in the limiting switching regimes $\nu\gg 1$ and $\nu\ll s$, this equation greatly simplifies.
In fact, using (\ref{eq:Nas}), equation (\ref{eq:beta}) becomes $(1+\beta/2)
\phi_{q(\beta)}^{(\infty,0)}=\phi_0^{(\infty,0)}$ 
when $\nu\gg 1$ and $\nu\ll s$, respectively.
Hence, the corresponding payoffs along $b=\beta(\nu,s)$ are $\Delta S_{\nu,\beta}=
\Delta F_{\nu,\beta}=(\phi_0^{(\infty)}-\phi_{\beta}^{(\infty)}){\cal K}\simeq
\beta \phi_{q(\beta)}^{(\infty)}{\cal K}/2$ when $\nu\gg 1$ and 
$\Delta S_{\nu,\beta}=\Delta F_{\nu,\beta}\simeq
\beta \phi_{q(\beta)}^{(0)}\langle K\rangle/2$ when  $\nu\ll s$, yielding 
$\Delta S_{\nu,\beta}/\langle N\rangle_{\nu,0}^*=\Delta F_{\nu,\beta}/\langle N\rangle_{\nu,0}^*\simeq
\beta \phi_{q(\beta)}^{(\infty,0)}/2$ in both limiting regimes.
It is however important to remember that in general $\phi_{\beta}\simeq \phi_{q(\beta)}$ depends nontrivially on $\nu$ and $s$, and can be either an increasing or decreasing function of $\nu$, see figures 2(a) and 3(a).

While the choice made here on how to measure the success of $S$ and $F$ is arguably the most natural, we could have also considered other
variants.
For instance, we could have considered that the best conditions to cooperate for the production of the public good would be: $S$ should receive a higher payoff than $F$, condition (a) as above, and, $S$ should  maximize the difference of payoffs $\Delta S_{\nu,b}- \Delta F_{\nu,b}$ (instead of condition (b)).
This would lead to an optimal value of the public good benefit $\widetilde{b}$ that would generally differ from $b^*$ especially at low switching rate (see figure 5(c)).
While this alternative definition of the optimal payoff for $S$ would lead to quantitative differences with the results reported in 
figure 6(b), the main qualitative features discussed here and in section IV.B would remain the same.

\section{
Effect of internal and environmental noise on population size distribution -- Linear noise approximation about the PDMP predictions}
\label{AppendixE} 
The PDMP approximation of the $N$-QSD can reproduce the number and location of its peaks, but fails to capture the width of the
distribution about the peaks and its accurate skewness, see, {\it e.g.}, figure \ref{fig:PDMP_joint}.
In this section, we derive the linear noise approximation (LNA) of the $N$-QSD used in Section V
to account for the demographic fluctuations about the PDMP predictions \cite{Hufton16}.

After the fixation of species $S$, $N_S=N$ and $N_F=0$, and the transition rates of the underpinning
birth-death process become $T_{S}^{+}=(1+b)N$,  $T_{S}^{-}=N^2/K(t)$, and $T_{F}^{\pm}=0$.
Similarly, after species $F$'s fixation, $N_F=N$ and $N_S=0$, and the transition rates (3) become $T_{F}^{+}=N$,  $T_{F}^{-}=N^2/K(t)$, with $T_{S}^{\pm}=0$.
To deal simultaneously with the ecological dynamics arising after the fixation of either species, it is convenient to define the 
auxiliary stochastic logistic process  $N \xrightarrow{T^{+}}  N + 1 \quad \text{and} \quad N  \xrightarrow{T^{-}}  N- 1$,
 with symmetric dichotomous Markov noise $\xi(t) \in \{-1,+1\}$ and randomly switching carrying capacity defined by equations (3) and (4).
This stochastic process is defined by the transition rates:
\begin{eqnarray}
 \label{eq:Taux}
 T^+=(1+q)N, \;\;
 T^-=\frac{N^2}{K(t)}=N^2\left[\frac{1}{{\cal K}} -\xi(t) \left(\frac{1}{{\cal K}} - \frac{1}{K_+}\right)\right],\;\;
\text{with}\;\; q=
\label{eq:qbis}
\begin{cases}
b &\mbox{after fixation of $S$}\\
0 &\mbox{after fixation of $F$}. 
 \end{cases}
 \end{eqnarray}
As explained in section V.B of the main text, it is convenient to work with the continuous  Markov 
process $\{n\equiv N/\Omega,\xi\}$ defined by
\begin{eqnarray}
 \label{eq:ntxit}
n \stackrel{{\cal T}^+}
     {\longrightarrow} n+\Omega^{-1}, \quad n \stackrel{{\cal T}^-}
     {\longrightarrow} n-\Omega^{-1}, \quad
\text{with}\; \xi \stackrel{\nu}{\longrightarrow} -\xi,
 \end{eqnarray}
with $ \Omega=\langle K \rangle$~\cite{comment2} and
\begin{eqnarray}
 \label{eq:kpm}
\psi\equiv\lim_{\Omega \to \infty} N/\Omega,
\quad
 \kappa\equiv{\cal K}/\Omega \quad  \text{and} \qquad
 \;
 	 k_{\xi}\equiv
	 \begin{cases}
 k_+= K_+/\Omega  &\mbox{if } \xi=1 \nonumber\\
 k_-=K_-/\Omega &\mbox{if } \xi=-1.\quad
 \end{cases}
\end{eqnarray}
Therefore, the transition rates for the process $\{n,\xi\}$ are
\begin{eqnarray}
 \label{eq:Tpm}
{\cal T}^{+}(\psi, \xi)=(1+q)\psi \quad \text{and} \quad
{\cal T}^{-} (\psi, \xi)=\psi^{2}\left\{\kappa^{-1}-\xi(\kappa^{-1}-k_+^{-1})
\right\}.
\end{eqnarray}
It is also useful to define $v_{\xi}$, associated with the deterministic flows of $\{n,\xi\}$, and $u_{\xi}$ associated with 
the diffusive flows:
\begin{eqnarray} 
\label{eq:vu}
v_{\xi} (\psi)\equiv {\cal T}^{+}-{\cal T}^{-}=\frac{{\cal F}(\Omega \psi,\xi)}{\Omega}
=\frac{\psi}{k_{\xi}}\left(\psi_{\xi}^*-\psi\right), \quad \text{and} \quad
u_{\xi}(\psi)\equiv {\cal T}^{+}+{\cal T}^{-}=\frac{\psi}{k_{\xi}}\left(\psi_{\xi}^*+\psi\right),
\end{eqnarray}
with $\psi_{\xi}^*=(1+q)k_{\xi}$.
It is worth noting that $v_{\xi}(\psi)>0$ when $\xi=+1$ and $v_{\xi}(\psi)<0$ when $\xi=-1$. 

When the environment is static ($K_{\pm}=K$), with $k_{\xi}=k$, $v_{\xi}=v$ and $u_{\xi}=u$, the LNA consists of performing a van Kampen
system size expansion of the underlying master equation, which yields the Fokker-Planck equation (FPE)
for the probability density $\pi(\eta,t)$~\cite{vanKampen,Gardiner}:
\begin{eqnarray}
 \label{eq:FPE1}
 \partial_t \pi(\eta,t)=- \partial_{\eta}\left[\eta v'(\psi)\pi(\eta,t)\right] +\frac{u}{2}\partial_{\eta}^2
 \pi(\eta,t),
\end{eqnarray}
where $v'=dv/d \psi$ and $\pi(\eta,t)$ is the PDF of the fluctuations $\{\eta(t)\}$ about the mean-field trajectory 
$\dot{\psi}=v(\psi)$.

Here, the environment varies stochastically by randomly switching between two states and we are interested
in the weak fluctuations about the PDMP trajectory $\psi(t)$. The process
$\{n,\xi\}$ is thus analyzed in terms of a ``pseudo-Fokker-Planck equation'' which
consists of an FPE, accounting for the internal noise, supplemented by terms arising from environmental stochasticity
via the PDMP $\{\psi(t)\}$ defined by the stochastic differential equation
\begin{eqnarray}
 \label{eq:LNA1}
\dot{\psi}=v_{\xi}(\psi)\,,
\end{eqnarray}
that is equivalent to (\ref{eq:sde}) and whose joint PDF 
is readily
obtained from (\ref{eq:pstarjoint}):
$\pi_{\nu,q}^*(\psi,\xi)= \Omega p^*(\Omega \psi, \xi)$.
Within the LNA, to account for the weak fluctuations about $\psi$ up to linear order in $\eta$, we 
obtain the following pseudo-FPE for the PDF $\pi_{\nu,q}(\psi, \eta, \xi,t)\equiv \pi(\psi,\eta, \xi)$ of the process (\ref{eq:Nflow}):
\begin{eqnarray}
 \label{eq:FPE_LNA}
 \partial_t \pi(\psi,\eta, \xi)&=&
 - \partial_{\eta}\left[\eta v_{\xi}'(\psi)\pi(\psi,\eta, \xi)\right]
+\frac{u_{\xi}}{2}\partial_{\eta}^2 \pi(\psi,\eta, \xi) \nonumber\\
&-&\partial_{\psi}\left[ v_{\xi}(\psi)\pi(\psi,\eta, \xi)\right]
-\nu\left[\pi(\psi, \eta, \xi)-\pi(\psi,\eta, -\xi)\right],
\end{eqnarray}
where, for notational simplicity, in this section we drop the time dependence
and the $\nu, q$ subscripts in the PDFs by writing $\pi(\eta, \xi)$ and $\pi(\psi, \eta, \xi)$ instead 
of $\pi_{\nu,q}(\eta, \xi,t)$ and $\pi_q(\psi, \eta, \xi,t)$,
etc.
On the RHS of eq.~(\ref{eq:FPE_LNA}), the first line corresponds to a usual FPE with a drift term $- \partial_{\eta}\left[\dots\right]$ and a diffusion coefficient $u_\xi$, while in the second line one recognizes the Liouvillian contribution $-\partial_{\psi}\left[ v_{\xi}(\psi)\pi(\psi,\eta, \xi)\right] $ and terms from random switching.

To determine the stationary Gaussian probability density $\pi^*(\eta| \psi, \xi)$ characterizing the demographic fluctuations 
$\eta$ about $\psi(t)$, we 
notice that $\pi(\psi,\eta,\xi)=\pi(\eta|\psi,\xi)\pi(\psi,\xi)$.
As explained in the main text, we then assume that demographic fluctuations about $\psi$ are the same
in each environmental state $\xi=\pm 1$, and write 
$\pi(\eta|\psi,\xi)=\pi(\eta|\psi)$~\cite{Hufton16}.
With this assumption, we can set $\partial_t (\pi^*(\eta, \psi, \xi)+
\pi^*(\eta, \psi, -\xi))=0$ and use equation (\ref{eq:FPE_LNA}) to obtain
\begin{eqnarray}
 \label{eq:PDF2_LNA}
 \hspace{-3mm}
 0= -\left[\pi^*(\xi|\psi)v_{\xi}'(\psi) + \pi^*(-\xi|\psi)v_{-\xi}'(\psi)  \right]\partial_{\eta}\left[\eta \pi^{*}(\eta|\psi)\right]
 +\frac{1}{2}\left[\pi^*(\xi|\psi) u_{\xi}(\psi)+\pi^*(-\xi|\psi) u_{-\xi}(\psi)\right]\partial_{\eta}^2 \pi^*(\eta|\psi),
\end{eqnarray}
where we have also used  $\pi^*(\psi,\xi)=\pi^*(\psi)\pi^*(\xi|\psi)$ and the zero-current boundary condition $\sum_{\xi} v_{\xi}\pi^*(\psi,\xi)=0$.
At the PDMP level, equation (\ref{eq:pcond}) expresses the probability of being in the environmental state $\xi$ given that the population has 
size $N$.
Hence, upon substituting $\pi^*(\xi|\psi)= -\xi v_{-\xi}/(\sum_{\xi=\pm 1} \xi v_{\xi})$, equation (\ref{eq:PDF2_LNA}) 
yields the stationary probability density $\pi^*(\eta|\psi)$ of an Ornstein-Uhlenbeck process \cite{Gardiner,vanKampen}.
In other words, $\pi^*(\eta|\psi)$ is a Gaussian with zero mean and variance
\begin{eqnarray}
 \label{eq:var}
\frac{u_{-}(\psi)v_{+}(\psi)-u_{+}(\psi)v_{-}(\psi)}
 {v_{-}(\psi)v_{+}'(\psi)-v_{+}(\psi)v_{-}'(\psi)}=\psi,
 \end{eqnarray}
where we have used equation (\ref{eq:vu}), and the subscripts $\pm$ refer to $\xi=\pm 1$. 
With equation (\ref{eq:var}), we find the stationary Gaussian probability density of the 
fluctuations  about $\psi$:
\begin{eqnarray}
 \label{eq:Gaussian}
 \pi^*(\eta|\psi)=\frac{e^{-\frac{\eta^2}{2\psi}}}{\sqrt{2\pi \psi}}.
 \end{eqnarray}

Within the LNA, see eq.~(25),
the marginal quasi-stationary PDF of the process $\{N(t),\xi(t)\}$ defined by (\ref{eq:Taux}) therefore is
\begin{eqnarray}
p_{{\rm LNA},\nu, q}^*(N)=\frac{\pi^*(n)}{\Omega}= \sum_{\xi=\pm 1} \int \int d\psi d\eta~\pi^*(\eta| \psi) \pi^*(\psi, \xi)~\delta\left(n-\psi-\frac{\eta} {\sqrt{\Omega}}\right)\,.\label{eq:PDF_LNA}
\end{eqnarray}

Upon substituting (\ref{eq:Gaussian}) and $\pi^*(\psi, \xi)=\Omega p_{\nu,q}^*(\Omega \psi,\xi)$ obtained from (\ref{eq:pstarjoint}), into (\ref{eq:PDF_LNA}),
we obtain the LNA-PDF of the process $\{N(t),\xi(t)\}$.
When $b=0$, the marginal LNA-PDF 
 $p_{{\rm LNA},0}^*(N)$ 
in the case of the pure resource competition is obtained from 
$\pi^*(n)$ with $q=0$ and reads
\begin{eqnarray} 
 \label{eq:PDF-LNAb0}
  \hspace{-8mm}
 p_{{\rm LNA},\nu,0}^*(N)  &\propto& 
\int \frac{d\eta~ e^{-\eta^2/[2(n-\eta/\sqrt{\Omega})]}}{\left(n-\frac{\eta}{\sqrt{\Omega}}\right)^{3/2}\left(k_{+}
-\left(n-\frac{\eta}{\sqrt{\Omega}}\right)\right)}~\left[\frac{\left\{k_{+} 
-\left(n-\frac{\eta}{\sqrt{\Omega}}\right)\right\}\left\{\left(n-\frac{\eta}{\sqrt{\Omega}}\right)- k_{-}\right\}}
{\left(n-\frac{\eta}{\sqrt{\Omega}}\right)^2}\right]^{\nu}\nonumber\\
&+& \int \frac{d\eta~ e^{-\eta^2/[2(n-\eta/\sqrt{\Omega})]}}{\left(n-\frac{\eta}{\sqrt{\Omega}}\right)^{3/2}\left( 
\left(n-\frac{\eta}{\sqrt{\Omega}}\right)-k_{-}\right)}~\left[\frac{\left\{k_{+} 
-\left(n-\frac{\eta}{\sqrt{\Omega}}\right)\right\}\left\{\left(n-\frac{\eta}{\sqrt{\Omega}}\right)-
k_-\right\}}{\left(n-\frac{\eta}{\sqrt{\Omega}}\right)^2}\right]^{\nu},
\end{eqnarray}
where $n=N/\Omega$, $k_{\pm}=K_{\pm}/\Omega$ and the proportional factor is the normalization constant.
In the public good scenario, $b>0$, the $F$-conditional LNA-PDF is  $p_{{\rm LNA},0}^*(N)$
while 
PDF conditioned on fixation of species $S$ (but unconditioned on $\xi$) is 
proportional to $\pi^*(n)$ with $q=b$, i.e. it is given by
\begin{eqnarray} 
 \label{eq:PDF-LNAb}
 \hspace{-8mm}
 p_{{\rm LNA},\nu,b}^*(N)   &\propto& \int \frac{d\eta~ e^{-\eta^2/[2(n-\eta/\sqrt{\Omega})]}}{\left(n-\frac{\eta}{\sqrt{\Omega}}\right)^{3/2}\left(\psi_{+}^*
-\left(n-\frac{\eta}{\sqrt{\Omega}}\right)\right)}~\left[\frac{\left\{\psi_{+}^* 
-\left(n-\frac{\eta}{\sqrt{\Omega}}\right)\right\}\left\{\left(n-\frac{\eta}{\sqrt{\Omega}}\right)- \psi_{-}^*\right\}}
{\left(n-\frac{\eta}{\sqrt{\Omega}}\right)^2}\right]^{\frac{\nu}{1+b}}\nonumber\\
&+& \int \frac{d\eta~ e^{-\eta^2/[2(n-\eta/\sqrt{\Omega})]}}{\left(n-\frac{\eta}{\sqrt{\Omega}}\right)^{3/2}\left( 
\left(n-\frac{\eta}{\sqrt{\Omega}}\right)-\psi_{-}^*\right)}~\left[\frac{\left\{\psi_{+}^* 
-\left(n-\frac{\eta}{\sqrt{\Omega}}\right)\right\}\left\{\left(n-\frac{\eta}{\sqrt{\Omega}}\right)-
\psi_{-}^*\right\}}{\left(n-\frac{\eta}{\sqrt{\Omega}}\right)^2}\right]^{\frac{\nu}{1+b}},
\end{eqnarray}
where  $\psi_{+}^*=(1+b)k_+$ and  $\psi_{-}^*=(1+b)k_-$.

\vspace{0.25cm}

The comparison between the LNA-PDFs and the $N-$QSD is shown in figures 7 and 8 of the main text, where a remarkable agreement is found when $b=0$ and $b>0$. 
However, as mentioned in the main text, some small deviations are observed  in figure 8(a),  at low  switching rate, near the peak of 
small intensity when $b>0$.
The possible reasons for these small deviations are multiple: When $\nu \ll 1$, the population near the peaks of weak intensity
is of size $N\approx (1+b)K_{\pm}$, and the assumption $\pi(\eta|\psi,\xi)\simeq \pi(\eta|\psi,-\xi)$ on which our LNA analysis is based may 
not be necessarily valid
since the fluctuations in the state $\xi=-1$ (with $N\approx (1+b)K_-$ and $b=2$) may be noticeably stronger than those in 
the state  $\xi=+1$ (where $N\approx (1+b)K_+$).
Furthermore, the peak in question is associated with the fixation of species $S$ for $b=2$
in a population of rather large size $\approx (1+b)K_+$, an event which occurs with a small probability that may be beyond the reach of the LNA. Moreover, 
the effective theory yielding the approximation $\phi\simeq \phi_q$ is based on the behavior at high switching rate and may be less accurate when 
$\nu \ll 1$ than in the regimes of intermediate and fast switching.

\end{widetext}


\begin{thebibliography}{99}
%

\bibitem{Morley83}
Morley C R, Trofymow J A, Coleman D C,  Cambardella C. 1983
Effects of freeze-thaw stress on bacterial populations in soil microcosms.
{\it Microbiol. Ecol.}  {\bf 9}, 329-340.
(doi: 10.1007/BF02019022)
%
\bibitem{Fux05}
Fux C A, Costerton J W, Stewart P S, Stoodley P. 2005 
Survival strategies of infectious biofilms.
{\it Trends Microbiol.} {\bf 13}, 34-40.
(doi: 10.1016/j.tim.2004.11.010)
 %
 \bibitem{May73}
May R M.1973
{\it Stability and complexity in model ecosystems}. Princeton, USA: Princeton University Press.
%
\bibitem{Karlin74}
Karlin S,  Levikson B. 1974
Temporal fluctuations in selection intensities: Case of small population size
{\it T.~Pop.~Biol.} {\bf 74}, 383-412.
(doi: 10.1016/0040-5809(74)90017-3)
%
\bibitem{Kussell05b}
Kussell E, Leibler S. 2005
Phenotypic Diversity, Population Growth, and Information in Fluctuating Environments
{\it Science} {\bf 309}, 2075-2078.
(doi: 10.1126/science.1114383)
%
%
\bibitem{Assaf13}
 Assaf M, Roberts E, Luthey-Schulten Z, Goldenfeld N. 2013 
 Extrinsic Noise Driven Phenotype Switching in a Self-Regulating Gene.
 {\it Phys.~Rev.~Lett.} {\bf 111}, 058102 (2013).
 (doi: 10.1103/PhysRevLett.111.058102)
%
\bibitem{He10}
He Q, Mobilia M, T\"auber U C. 2010 
Spatial rock-paper-scissors models with inhomogeneous reaction rates.
{\it Phys.~Rev.~E} {\bf 82}, 051909.
(doi: 10.1103/PhysRevE.82.051909)
%
 \bibitem{Tauber13}
Dobramysl U,  T\"auber U C. 2013
Environmental Versus Demographic Variability in Two-Species Predator-Prey Models.
{\it Phys. Rev. Lett.} {\bf 110}, 048105.
(doi: 10.1103/PhysRevLett.110.048105)
%
\bibitem{AMR13}
Assaf M, Mobilia M, Roberts E. 2013
Cooperation Dilemma in Finite Populations under Fluctuating Environments.
{\it Phys.~Rev.~Lett.} {\bf 111}, 238101.
(doi: 10.1103/PhysRevLett.111.238101)
%
\bibitem{Ashcroft14}
Ashcroft P, Altrock P M, Galla T. 2014 
Fixation in finite populations evolving in fluctuating environments.
{\it J. R. Soc. Interface} {\bf 11}, 20140663.
(doi: 10.1098/rsif.2014.0663 )
%
\bibitem{Melbinger15}
Melbinger A, Vergassola M. 2015 
The Impact of Environmental Fluctuations on Evolutionary Fitness Functions
{\it Scientific Reports} {\bf 5}, 15211.
(doi: 10.1038/srep15211)
%
%
\bibitem{Hufton16}
Hufton P G, Lin Y T, Galla T, McKane A J. 2016
Intrinsic noise in systems with switching environments.
{\it Phys.~Rev.~E} {\bf 93}, 052119.
(doi: 10.1103/PhysRevE.93.052119)
%
\bibitem{Hidalgo17}
Hidalgo J, Suweis S, Maritan A. 2017
Species coexistence in a neutral dynamics with environmental noise.
{\it J.~Theor.~Biol.} {\bf 413}, 1-10.
(doi: 10.1016/j.jtbi.2016.11.002)
%
\bibitem{Shnerb17}
Danino M, Shnerb N M. 2018
Fixation and absorption in a  fluctuating environment.
{\it J.~Theor.~Biol.} {\bf 441}, 84-92.
(doi: 10.1016/j.jtbi.2018.01.004)
%
%
\bibitem{Chesson81}
Chesson P L,  Warner R R. 1981
Environmental Variability Promotes Coexistence in Lottery Competitive Systems.
{\it American Naturalist} {\bf 117}, 923-943.
(doi: 10.1086/283778)
%
\bibitem{Kussell05}
Kussell E, Kishony R,  Balaban N Q,  Leibler S. 2005
Bacterial Persistence: A Model of Survival in Changing Environments.
{\it Genetics} {\bf 169}, 1807-1814.
(doi: 10.1534/genetics.104.035352)
%
\bibitem{Acar08}
Acer M, Mettetal J,  van Oudenaarden A. 2008 
Stochastic switching as a survival strategy in fluctuating environments.
{\it Nature Genetics} {\bf 40}, 471-475.
(doi: 10.1038/ng.110)
%
%
\bibitem{Loreau08}
Loreau, M,  de Mazancourt C. 2008
Species synchrony and its drivers: neutral and noneutral community dynamics in fluctuating environments
{\it American Naturalist} {\bf 172}, E49-E66.
(doi:10.1086/589746)
%
\bibitem{Beaumont09}
Beaumont H, Gallie J, Kost C, Ferguson G,  Rainey P. 2009
Experimental evolution of bet hedging 
Nature  {\bf 462}, 90-93.
(doi: 10.1038/nature08504)
%
\bibitem{Visco10}
Visco P, Allen R J,  Majumdar S N, Evans M R. 2010
Switching and Growth for Microbial Populations in Catastrophic Responsive Environments.
{\it Biophys. J.} {\bf 98}, 1099-1108.
(doi: 10.1016/j.bpj.2009.11.049)
%
\bibitem{Xue17}
Xue B, Leibler S. 2017
Bet-hedging against demographic fluctuations
{\it Phys.~Rev.~Lett.} {\bf 119}, 108113.
 (doi: 10.1103/PhysRevLett.111.058102)
%
\bibitem{Kimura} 
Crow J F, Kimura M, 2009
{\it An Introduction to Population Genetics Theory}.
New Jersey, USA: Blackburn Press
%


\bibitem{Blythe07}
Blythe R A,  McKane A J. 2007
Stochastic models of evolution in genetics, ecology and linguistics
{\it J. Stat. Mech.} {\bf P07018}
(doi:10.1088/17442-5468/2007/07/P07018)
%

\bibitem{Nowak}
Nowak M A, 2006
{\it Evolutionary
Dynamics}. Cambridge, USA: Belknap Press
 

 
\bibitem{Roughgarden79}
Roughgarden J 1979 {\it Theory of Population Genetics and Evolutionary Ecology: An Introduction}.
New York, USA: Macmillan.
%
\bibitem{Melbinger2010}
 Melbinger A, Cremer J, Frey E. 2010
 Evolutionary Game Theory in Growing Populations.
{\it   Phys.~Rev.~Lett.} {\bf 105}, 178101.
(doi:10.1103/PhysRevLett.105.178101)
%
 \bibitem{Cremer2011} 
Cremer J, Melbinger A,  Frey E. 2011  
 Evolutionary and population dynamics: A coupled approach.
{\it  Phys.~Rev.~E} {\bf 84}, 051921. (doi: 10.1103/PhysRevE.84.051921)
%
 \bibitem{Pelletier09} 
Pelletier F, Garant D, Hendry H P. 2009 Eco-evolutionary dynamics.
{\it Phil. Trans. R. Soc. B} {\bf 364} 1483-1489. (doi:10.1098/rstb.2009.0027)
%
 \bibitem{Harrington14} 
Harrington K I, Sanchez A. 2014 Eco-evolutionary dynamics of complex strategies in microbial communities.
{\it Communicative \& Integrative Biology} {\bf 7:1}, e28230:1-7. (doi:10.4161/cib.28230)
%
%
\bibitem{Leibler09}
Chuang J S,  Rivoire O,  Leibler S. 2009
Simpson's Paradox in a Synthetic Microbial System.
{\it Science} {\bf 323}, 272-275.
(doi: 10.1126/science.1166739)
%
 \bibitem{Melbinger2015a}
 Melbinger A, Cremer J, Frey E. 2015 
 The emergence of cooperation from a single mutant during microbial life cycles.
{ J.~R.~Soc. Interface} {\bf 12}, 20150171.
(doi: 10.1098/rsif.2015.0171)
%

\bibitem{Wahl02}
  Wahl L M, Gerrish P J,  Saika-Voivod I. 2002 
 Evaluating the impact of population bottlenecks in experimental evolution.
 {\it Genetics} {\bf 162}, 961-971.
 (url: http://www.genetics.org/content/162/2/961)
%
\bibitem{Patwas09}
Patwas Z, Wahl L M. 2009
Adaptation rates of lytic viruses depend critically on whether host cells survive the bottleneck.
{\it Evolution} {\bf 64}, 1166-1172.
(doi: 10.1111/j.1558-5646.2009.00887.x)
%
\bibitem{Brockhurst07a}
Brockhurst M A, Buckling A, Gardner A. 2007
Cooperation Peaks at Intermediate Disturbance.
{\it Curr. Biol.} {\bf 17}, 761-765.
(doi: 10.1016/j.cub.2007.02.057)
%
\bibitem{Brockhurst07b}
Brockhurst M A. 2007
Population Bottlenecks Promote Cooperation in Bacterial Biofilms.
{PLoS One} {\bf 7}, e634.
(doi: 10.1371/journal.pone.0000634)
%

\bibitem{Wienand15}
Wienand K, Lechner M,  Becker F,  Jung H,  Frey E. 2015 
Non-Selective Evolution of Growing Populations.
PloS one, {\bf 10(8)}, e0134300.
(doi: 10.1371/journal.pone.0134300)
%
\bibitem{Wienand18}
Becker F, Wienand K, Lechner M, Frey E, Jung H. 2018
Interactions mediated by a public good transiently increase cooperativity in growing Pseudomonas putida 
metapopulations.
Scientific Reports, {\bf 8}, 4093.
(doi: 10.1038/s41598-018-22306-9)
%

\bibitem{Rainey03}
Rainey P B, Rainey K. 2003  
Evolution of cooperation and conflict in experimental bacterial populations
{\it Nature} {\bf 425}, 72.
(doi: 10.1038/nature01906)
%

\bibitem{KEM1}
Wienand K, Frey E, Mobilia M. 2017 
Evolution of a Fluctuating Population in a Randomly Switching Environment.
{\it Phys. Rev. Lett.} {\bf 119}, 158301. (doi:10.1103/PhysRevLett.119.158301)
%
\bibitem{Supp}
Wienand K, Frey E, Mobilia M. 2018
{\it Supplementary Material} (SM). This SM is provided as an appendix to this main document (see pages 13-25).
The SM along with additional resources are electronically available at
{\it fighshare https://doi.org/10.6084/m9.figshare.5683762}.
(doi:10.6084/m9.figshare.5683762)
%
\bibitem{Cremer09}
Cremer J, Reichenbach T, Frey E. 2009
The edge of neutral evolution in social dilemmas
{\it New J. Phys.} {\bf 11}, 093029.
(doi: 10.1088/1367-2630/11/9/093029)
%
\bibitem{Strains}
Here, in reference to 
microbial communities, we use interchangeably use the terms ``species'' and ``strain''.
%
\bibitem{HL06}
Horsthemke W, Lefever R. 2006
{\it Noise-Induced Transitions}. Berlin, Germany: Springer
%
%
\bibitem{Bena06}
Bena I. 2006 
Dichotomous Markov noise: Exact results for out-of-equilibrium systems. A review.
{\it Int.~J.~Mod.~Phys. B} {\bf 20}, 2825-2889.
(doi: 10.1142/S0217979206034881)
%

\bibitem{Gardiner} 
Gardiner C W. 2002
{\it Handbook of Stochastic Methods}
New York, USA: Springer
%
%
\bibitem{vanKampen} 
van Kampen N G. 2003
{\it Stochastic Processes in Physics and Chemistry}
Amsterdam, The Netherlands: North-Holland Publishing
%


\bibitem{Kithara79}
Kitahara K, Horsthemke W, Lefever R. 1979
Coloured-noise-induced transitions: exact results for external dichotomous Markovian noise.
{\it Phys.~Lett.} {\bf 70A}, 377-380.
(doi: 10.1016/0375-9601(79)90336-0)

\bibitem{Hanggi81}
H\"anggi P, Talkner P. 1981
Non-Markov processes: The problem of the mean first passage time.
{\it Z.~Phys. B} {\bf 45}, 79-83.
 (doi: 10.1007/BF01294279)

\bibitem{Broek84}
Van den Broek C,  H\"anggi P. 1984
Activation rates for nonlinear stochastic flows driven by non-Gaussian noise.
{\it Phys. Rev. A} {\bf 30}, 2730-2736. 
(doi: 10.1103/PhysRevA.30.2730)

\bibitem{Sancho85}
Sancho J M. 1985 
External dichotomous noise: The problem of the mean-first-passage time.
{\it Phys. Rev. A} {\bf 31}, 3523(R)-3526(R).
(doi: 10.1103/PhysRevA.31.3523)

\bibitem{Spalding17}
Spalding C, Doering C  R, Flierl G R. 2017
Resonant activation of population extinctions.
{\it Phys. Rev. E} {\bf 96}, 042411. 
(doi: 10.1103/PhysRevE.96.042411)
%
\bibitem{meta}
A finite population unavoidably collapses into  $(N,x)=(0,0)$ where it is extinct~\cite{Spalding17}.
This phenomenon is practically unobservable when $K_-\gg 1$ and occurs after  lingering in the system's quasi-stationary state (where $N$ is distributed according to its $N$-QSD) and much after the fixation of one species.



\bibitem{SM}
Wienand K, Frey E, Mobilia M. 2017
{\it 
http://link.aps.org/supplemental/10.1103/PhysRevLett.119.158301}.
(doi: 10.1103/PhysRevLett.119.158301)
%


\bibitem{Cremer17}
Cremer J, Arnoldini M, Hwa T. 2017
Effect of water flow and chemical environment on microbiota growth and
composition in the human colon.
{\it Proc. Nat. Acad. Sci.} \textbf{25}, 6438-6443.
(doi: 10.1073/pnas.1619598114)

\bibitem{Lindsey13}
Lindsey H A, Gallie J, Taylor S, Kerr B. 2013
Evolutionary rescue from extinction is contingent on a lower rate of
environmental change.
{\it Nature} \textbf{494}, 463-467.
(doi: 10.1038/nature11879)

\bibitem{Lambert15}
Lambert G, Kussell E. 2015
Quantifying Selective Pressures Driving Bacterial Evolution Using
Lineage Analysis.
{\it Phys. Rev. X} \textbf{5}, 011016.
(doi: 10.1103/PhysRevX.5.011016)

\bibitem{Lechner16}
Lechner M, Schwarz M, Opitz M, Frey E. 2016
Hierarchical Post-transcriptional Regulation of Colicin E2 Expression
in \textit{Escherichia coli}.
{\it PLOS Comp. Biol.} \textbf{12}, e1005243.
(doi: 10.1371/journal.pcbi.1005243)



\bibitem{Davis84}
Davis M H A. 1984
Piecewise--deterministic Markov processes: a general class of non–diffusion stochastic models
{\it J.~R. Stat. Soc. B} {\bf 46}, 353-388.
(Retrieved from http://www.jstor.org/stable/2345677)
%
%

%
\bibitem{Moran}
Moran P A P. 1962  {\it The statistical processes of evolutionary theory}. 
Oxford, UK: Clarendon

%

\bibitem{Antal06}
Antal I, Scheuring I. 2006 
Fixation of Strategies for an Evolutionary Game in Finite Populations
{\it Bull. Math. Biol.} {\bf 68}, 1923-1944.
(doi: 10.1007/s11538-006-9061-4)
%
\bibitem{Otto97}
Otto S P, Whitlock M C. 1997
The Probability of Fixation in Populations of Changing Size.
{\it Genetics} \textbf{146}, 723-733.
(url: http://www.genetics.org/content/146/2/723)


\bibitem{videos}
Wienand K, Frey E, Mobilia M. 2017 
{\it fighshare https://doi.org/10.6084/m9.figshare.5082712}.
(doi:10.6084/m9.figshare.5082712.v5)
%

%
%
\bibitem{Mathematica}
Wolfram Research. 2010 
Mathematica, Version 10.0
{\it Wolfram Research Inc}
%
%
\bibitem{Gillespie76} 
Gillespie D T. 1976
A general method for numerically simulating the stochastic time evolution of coupled chemical reactions.
{\it J. Comput. Phys.} \textbf{22}, 403.
(doi: 10.1016/0021-9991(76)90041-3)
%

%
\bibitem{Ewens} 
 Ewens E W, 2004
 {\it Mathematical Population Genetics}. 
 New York, USA: Springer.



\bibitem{comment}
With eq.~(\ref{eq:pstarqmar}), in the realm of the PDMP approximation, we find that $(1+b\phi_b) \langle N \rangle_{0,\nu}^*\leq \langle N \rangle_{b,\nu}^*\leq (1+b\phi_b)\langle K \rangle$.

\bibitem{Broom13}
 Broom M, Rycht\'a\v{r} J. 2013.
 {\it Game-Theoretical Models in Biology}.
 Boca Raton, USA: CRC Press
	%


\bibitem{Gardiner} 
Gardiner C W. 2002
{\it Handbook of Stochastic Methods}
New York, USA: Springer
%

%
\bibitem{vanKampen} 
van Kampen N G. 2003
{\it Stochastic Processes in Physics and Chemistry}
Amsterdam, The Netherlands: North-Holland Publishing
%

 
 
\bibitem{comment2}
Since $K_{\pm}$ are here assumed to be of the same order, with $K_+\gtrsim K_-\gg 1$
with $ \Omega=\langle K \rangle$~\cite{comment2}, we 
could also define $\Omega=K_+$ or $\Omega= K_-$ and proceed similarly.


%
\end{thebibliography}
\end{document}